\documentclass[twocolumn,floatfix,prb,aps,showpacs]{revtex4-1}
\usepackage{graphicx,amsmath,amssymb,color,xcolor}
\usepackage{nicefrac}
\usepackage[titletoc,title]{appendix}

\newcommand{\be}{\begin{equation}}
\newcommand{\ee}{\end{equation}}

\newcommand{\ba}{\begin{eqnarray}}
\newcommand{\ea}{\end{eqnarray}}

\begin{document}

\title{Fractional quantum Hall effect at $\nu=2+4/9$}
\author{Ajit C. Balram$^{1}$ and A. W\'ojs$^{2}$}
\affiliation{$^{1}$Institute of Mathematical Sciences, HBNI, CIT Campus, Chennai 600113, India}
\affiliation{$^{2}$Department of Theoretical Physics, Wroc\l{}aw University of Science and Technology, 50-370 Wroc\l{}aw, Poland}

\date{\today}

\begin{abstract} 
Motivated by two independent experiments revealing a resistance minimum at the Landau level (LL) filling factor $\nu=2+4/9$, characteristic of the fractional quantum Hall effect (FQHE) and suggesting electron condensation into a yet unknown quantum liquid, we propose that this state likely belongs in a parton sequence, put forth recently to understand the emergence of FQHE at $\nu=2+6/13$. While the $\nu=2+4/9$ state proposed here directly follows three simpler parton states, all known to occur in the second LL, it is topologically distinct from the Jain composite fermion (CF) state which occurs at the same $\nu=4/9$ filling of the lowest LL. We predict experimentally measurable properties of the $4/9$ parton state that can reveal its underlying topological structure and definitively distinguish it from the $4/9$ Jain CF state. 
\end{abstract}
\maketitle

The fractional quantum Hall effect (FQHE)~\cite{Tsui82,Laughlin83} forms a paradigm in our understanding of strongly correlated quantum phases of matter. Of particular interest among the panoply of FQHE phases are the ones observed in the second Landau level (SLL) of ordinary semiconductors such as GaAs. These have attracted widespread attention because of the possibility that the excitations of these phases obey non-Abelian braiding statistics, which could potentially be utilized in carrying out fault-tolerant topological quantum computation~\cite{Kitaev03,Nayak08}. 

FQHE was first observed at filling factor $\nu=1/3$~\cite{Tsui82} in the lowest LL (LLL) and was explained by Laughlin using his eponymous wave function~\cite{Laughlin83}. Soon a whole zoo of fractions were observed, primarily along the sequence $n/(2pn\pm 1)$ ($n$ and $p$ are positive integers) and its particle-hole conjugate~\cite{Eisenstein90}. These FQHE states can be understood as arising from the integer quantum Hall effect (IQHE) of composite fermions (CFs)~\cite{Jain89,Jain07}, which are bounds states of electron and an even number ($2p$) of quantized vortices. The theory of weakly interacting CFs captures almost all of the observed FQHE phenomenology in the LLL.

In comparison to the FQHE in the LLL, the FQH states in the SLL are fewer in number and are more fragile~\cite{Willett87}. Moreover, the nature of many FQH states in the SLL is dramatically different from their LLL counterparts. In particular, one of the strongest FQH states in the SLL occurs at $\nu=2+1/2$~\cite{Willett87}, whereas at the corresponding half-filled LLL a compressible state is observed. A key breakthrough in the field of FQHE came about by a proposal of Moore and Read~\cite{Moore91}, who posited a ``Pfaffian'' wave function to describe the state at $\nu=5/2$. Subsequently, it was understood that the Pfaffian wave function can be interpreted as a $p$-wave paired state of composite fermions~\cite{Read00}. The excitations of the Pfaffian state are Majorana fermions which, owing to their non-Abelian braiding properties, could form building blocks of a topological quantum computer~\cite{Kitaev03,Nayak08}.

The nature of the state at $1/3$ state in the SLL, although believed to be Laughlin-like, has been under intense debate~\cite{Ambrumenil88,Balram13b,Johri14,Peterson15,Kleinbaum15,Jeong16,Balram19a}. FQHE has been observed at $\nu = 2 + 2/5$~\cite{Xia04,Choi08,Pan08,Kumar10} but is widely believed to be of a parafermionic nature, unlike the Abelian LLL Jain CF state at $\nu=2/5$~\cite{Read99,Rezayi09,Wojs09,Bonderson12,Sreejith13,Zhu15,Mong15,Pakrouski16}. Furthermore, as yet there is no conclusive experimental evidence of FQHE at the next three members of the $n/(2n+1)$ Jain sequence, namely $\nu = 3/7$, $4/9$, and $5/11$, in the SLL~\cite{Shingla18}, though some features of FQHE have been reported in the literature at some of these fillings~\cite{Pan08,Choi08,Reichl14} (A non-Abelian parton state at $3/7$ was constructed in Ref.~\cite{Faugno20a} and shown to be feasible in the SLL.). However, FQHE has been observed at $\nu=2+6/13$~\cite{Kumar10}. These observations collectively point to the fact that the nature of FQHE states in the LLL and SLL are different from each other, and a description of FQHE states in the SLL likely entails going beyond the framework of noninteracting composite fermions. 

Recently, candidate ``parton'' states have been constructed to describe the FQHE at many of the experimentally observed fillings in the SLL~\cite{Balram18,Balram18a,Balram19}. The parton theory~\cite{Jain89b} produces model incompressible states beyond the Laughlin~\cite{Laughlin83} and the CF theory~\cite{Jain89}. The ``$\bar{n}\bar{2}111$'' parton states for $n=1,2,3$ give a good description of the SLL FQHE states observed at $\nu=2+2/3,2+1/2$ and $2+6/13$~\cite{Balram18,Balram18a}. In particular, the $\bar{n}\bar{2}111$ sequence underscores the unusual stability of $6/13$ in the SLL.

In this work, we consider the next member of the $\bar{n}\bar{2}111$ parton sequence, namely $\bar{4}\bar{2}111$ and consider its feasibility at $\nu=2+4/9$. While FQHE at $2+4/9$ has not been established conclusively, indications for it in the form of a minimum in the longitudinal resistance have already been seen in experiments~\cite{Choi08,Reichl14}. Future experiments on high-quality samples could likely establish FQHE at this filling by an observation of a well-quantized plateau in the Hall resistance. We show that the $\bar{4}\bar{2}111$ parton state gives a good description of the exact SLL Coulomb ground state seen in numerics. Furthermore, the parton state is energetically favorable compared to the  $411$ Jain CF state at $2+4/9$ in the thermodynamic limit. Therefore, if FQHE is established at $\nu=2+4/9$, it is highly likely to be distinct from its LLL counterpart at $\nu=4/9$, which is well described by the $411$ Jain CF state. The parton state is topologically distinct from the Jain CF state and we make predictions for experimentally measurable quantities that can unambiguously distinguish the two. In particular, the parton state supports counter-propagating edge modes that do not occur in the Jain CF state. 

The parton theory put forth by Jain~\cite{Jain89b} constructs FQHE states as a product of IQH states. The essential idea is to break each electron into fictitious objects called partons, place the partons into incompressible IQH states, and recover the final state by fusing the partons back into the physical electrons. The $N$-electron wave function for the $l$-parton Jain state, denoted as ``$n_{1}n_{2}\cdots n_l$'', is given by
\begin{equation}
\label{eq:general_parton}
\Psi^{n_{1}n_{2}\cdots n_l}_{\nu}= \mathcal{P}_{\rm LLL} \prod_{\mu=1}^l \Phi_{n_\mu}(\{z_j\}). 
\end{equation}
Here $z_j = x_j - i y_j$ is the two-dimensional coordinate of electron $j$ which is parametrized as a complex number, $\mu$ denotes the parton species and $\mathcal{P}_{\rm LLL}$ implements projection into the LLL. Each parton species is exposed to the external magnetic field and occupies the same area, which fixes the charge of the $\mu$ parton species to $q_\mu=-e\nu/n_\mu$, with $\sum_\mu q_\mu=-e$, where $-e$ is the charge of the electron. The state $\Phi_{n_\mu}$ is the Slater determinant IQH wave function of $N$ electrons filling the lowest $|n_{\mu}|$ LLs. We allow for $n_{\mu}<0$ and negative values are denoted by $\bar{n}$, with $\Phi_{\bar{n}}=\Phi_{-|n|}=[\Phi_n]^*$. The parton state of Eq.~(\ref{eq:general_parton}) occurs at a filling factor of $\nu=( \sum_\mu n_\mu^{-1} )^{-1}$ and has a shift~\cite{Wen92} of $\mathcal{S}=\sum_\mu n_\mu$ in the spherical geometry. 

The Laughlin state~\cite{Laughlin83} is a ``$11\cdots$'' parton state. The parton state $n11$ ($\bar{n}11$) with wave function $\Psi^{\rm CF}_{n/(2n+1)} = \mathcal{P}_{\rm LLL}  \Phi_{n}\Phi^{2}_{1}$ ($\Psi^{\rm CF}_{n/(2n-1)} = \mathcal{P}_{\rm LLL}  [\Phi_{n}]^{*}\Phi^{2}_{1}$) correspond to the Jain CF states~\cite{Jain89}. Recently, it has been shown that parton states of the form ``$221\cdots,$'' which are not composite fermion states, could be viable candidates to describe certain FQH states observed in the LLL in wide quantum wells~\cite{Faugno19} and in LLs of graphene~\cite{Wu16,Kim18}.

The motivation for considering the $\bar{4}\bar{2}111$ parton state stems from the recent application of parton theory to capture states in the SLL~\cite{Balram18,Balram18a,Balram19}. Consider the family of parton states described by the wave function
\begin{equation}
\Psi^{\bar{n}\bar{2}111}_{\nu=2n/(5n-2)}=\mathcal{P}_{\rm LLL}  \Phi_{\bar{n}}\Phi_{\bar{2}}\Phi_1^3 \sim \frac{\Psi^{\rm CF}_{n/(2n-1)}\Psi^{\rm CF}_{2/3}}{\Phi_{1}}.
\label{eq:parton_barnbar2111} 
\end{equation}
The $\sim$ sign in Eq.~(\ref{eq:parton_barnbar2111}) indicates that the states written at both sides of the sign differ slightly in the details of the projection. We expect that such details do not change the topological nature of the states~\cite{Balram16b}. Throughout this text, for the $\bar{n}\bar{2}111$ parton states we use the form given on the rightmost side of Eq.~(\ref{eq:parton_barnbar2111}). A nice feature of the parton wave functions stated in Eq.~(\ref{eq:parton_barnbar2111}) is that they can be evaluated for large system sizes which allow a reliable extrapolation of their thermodynamic energies. One can construct the above parton states for large system sizes because the constituent Jain CF states can be evaluated for hundreds of electrons using the Jain-Kamilla method of projection~\cite{Jain97b,Moller05,Jain07,Davenport12,Balram15a}. The Jain CF states in this work are evaluated using the Jain-Kamilla method.

The $n=1$ member, namely $\bar{1}\bar{2}111$, is likely topologically equivalent to the $\nu=2/3$ $\bar{2}11$ Jain CF state~\cite{Balram16b}. The $n=2$ member $\bar{2}\bar{2}111$ has a good overlap with the exact SLL Coulomb ground state at $\nu=5/2$~\cite{Balram18}. The $n=3$ state, $\bar{3}\bar{2}111$, gives a good description of the Coulomb ground state at $\nu=2+6/13$~\cite{Balram18a}. In the Supplemental Material (SM)~\cite{SM} we provide further evidence in favor of the feasibility of the $\bar{3}\bar{2}111$ parton state to describe the $\nu=2+6/13$ FQHE. We shall consider the $n=4$ member of this sequence which occurs at filling factor $4/9$.

Although there is no definitive observation of FQHE at $4/9$ in the SLL, signatures of incompressibility have been seen at $\nu=2+4/9$ and its particle-hole conjugate at $\nu=2+5/9$~\cite{Choi08,Reichl14}. FQHE at $\nu=2+4/9$ is likely swamped by a bubble phase~\cite{Shingla18}; however, it is likely that with improvements in the sample quality or for some interaction parameters close to that of the SLL Coulomb one, FQHE will ultimately be observed at $2+4/9$. 

For all our calculations, we deploy Haldane's spherical geometry~\cite{Haldane83}, in which $N$ electrons reside on the surface of a sphere in the presence of a radial magnetic field generated by a monopole of strength $2Q (hc/e)$ located at the center of the sphere. FQHE ground states occur at flux values $2Q=\nu^{-1}N-\mathcal{S}$, where $\mathcal{S}$ is a rational number called the shift, which is useful in characterizing the topological nature of the FQHE state~\cite{Wen92}. All FQHE ground states are uniform on the sphere and thus have total orbital angular momentum $L=0$. The parton states $\Psi^{\bar{n}\bar{2}111}_{\nu=2n/(5n-2)}$ of Eq.~(\ref{eq:parton_barnbar2111}) satisfy the flux-particle relationship $2Q = [(5n-2)/(2n)]N - (1-n)$; i.e., their filling factors are $\nu=2n/(5n-2)$ and their shifts are $\mathcal{S}=1-n$. Of particular interest to us in this work is the $\bar{4}\bar{2}111$ parton state which has a shift of $\mathcal{S}=-3$. This parton state is topologically distinct from the $411$ Jain CF state which also occurs at $\nu=4/9$ but has a shift of $\mathcal{S}=6$. We assume a single-component system and neglect the effects of LL mixing and disorder. Under these assumptions, states related by particle-hole conjugation are considered on the same footing.

Throughout this work, we shall write wave functions in the LLL, which is where they are easily evaluated, even though they might apply to states occurring in the SLL. Haldane~\cite{Haldane83} showed that the physics of the SLL can be simulated in the LLL by using an effective interaction that has the same set of Haldane pseudopotentials in the LLL as the Coulomb interaction has in the SLL. In this work, we have used the form of the effective interaction described in Ref.~\cite{Shi08} to simulate the physics of the SLL in the LLL. 

\begin{figure}[t]
\begin{center}
\includegraphics[width=0.47\textwidth,height=0.23\textwidth]{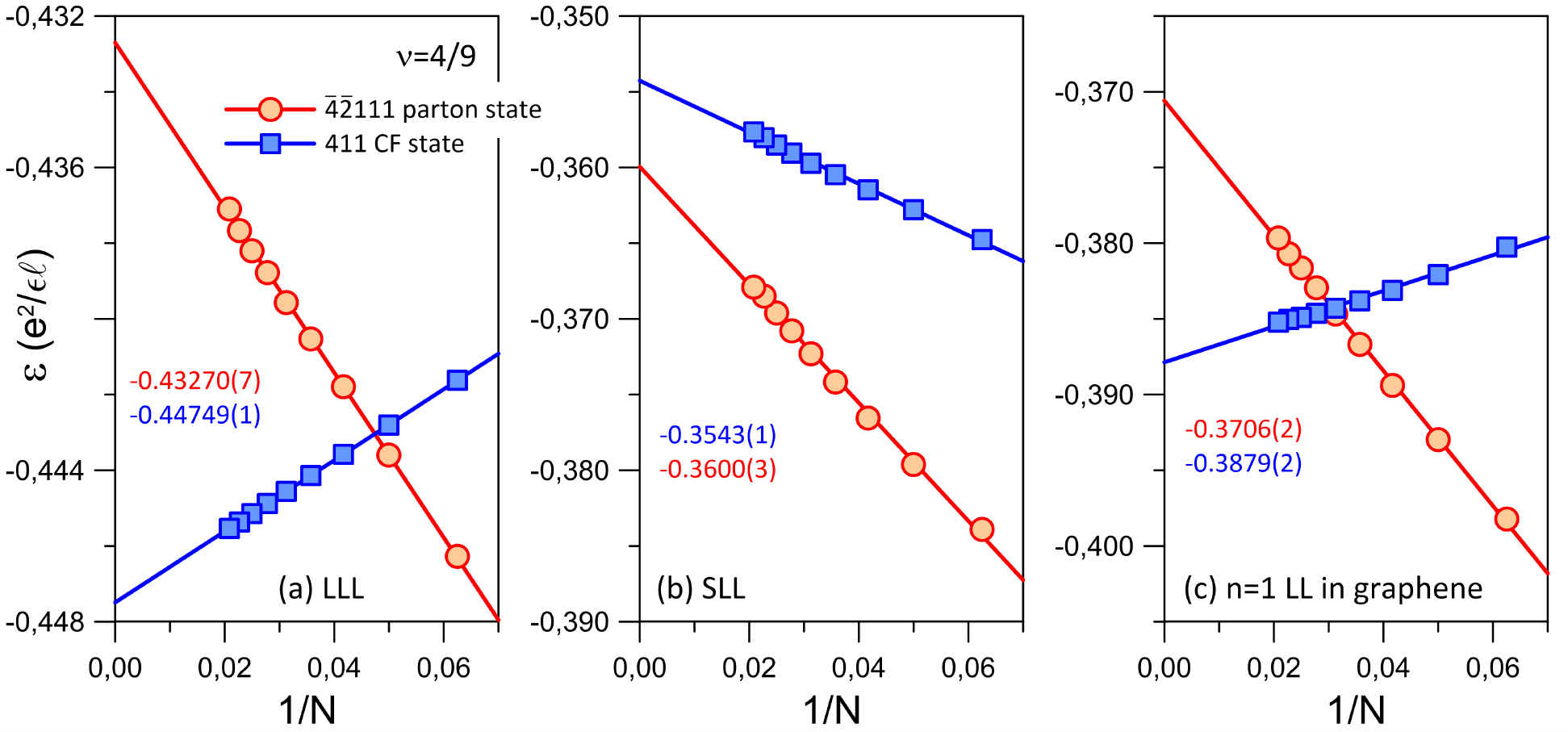} 
\caption{(color online) Thermodynamic extrapolations of the Coulomb per-particle energies for the $411$ Jain composite fermion (CF) state and the $\bar{4}\bar{2}111$ parton state. The left-hand, middle, and right-hand panels show energies for $\nu=4/9$ in the $n=0$ LL, $n=1$ LL of GaAs, and in the $n=1$ LL of monolayer graphene respectively. The extrapolated energies, obtained from a linear fit in $1/N$, are quoted in units of $e^2/(\epsilon\ell)$ on the plot (error in the fit is indicated in the parentheses). The energies include contributions of electron-background and background-background interactions and are density corrected~\cite{Morf86}. The LLL Coulomb energy for the $\nu=4/9$ Jain CF state has been reproduced from Ref.~\cite{Balram17}. }
\label{fig:extrapolations_energies_4_9}
\end{center}
\end{figure}

Let us begin by testing the viability of the $\bar{4}\bar{2}111$ parton state for $\nu = 2 + 4/9$ FQHE. In Fig.~\ref{fig:extrapolations_energies_4_9} we compare the energies of the $\bar{4}\bar{2}111$ parton and the $411$ Jain CF states at $\nu=4/9$ in the LLL and the SLL. In the LLL, as expected, we find that the Jain CF state has lower energy than the parton state. However, in the SLL we find that the $\bar{4}\bar{2}111$ parton state is energetically more favorable compared to the Jain CF state. For the sake of completeness, we have also investigated the competition between the parton and Jain CF states in the $n=1$ LL of monolayer graphene. The effective interaction we use to simulate the physics of the $n=1$ LL of monolayer graphene in the LLL is described in Ref.~\cite{Balram15c}. We find the $411$ Jain CF state has lower energy here, consistent with the fact that experimentally observed FQHE states in the $n=1$ LL of monolayer graphene are well described by the CF paradigm~\cite{Amet15,Balram15c,Zeng18}. Results for $n=0$ LL of graphene are identical to those in the LLL of GaAs under our working assumptions of neglecting effects of finite width and LL mixing.

\begin{figure}[t]
\begin{center}
\includegraphics[width=0.37\textwidth,height=0.29\textwidth]{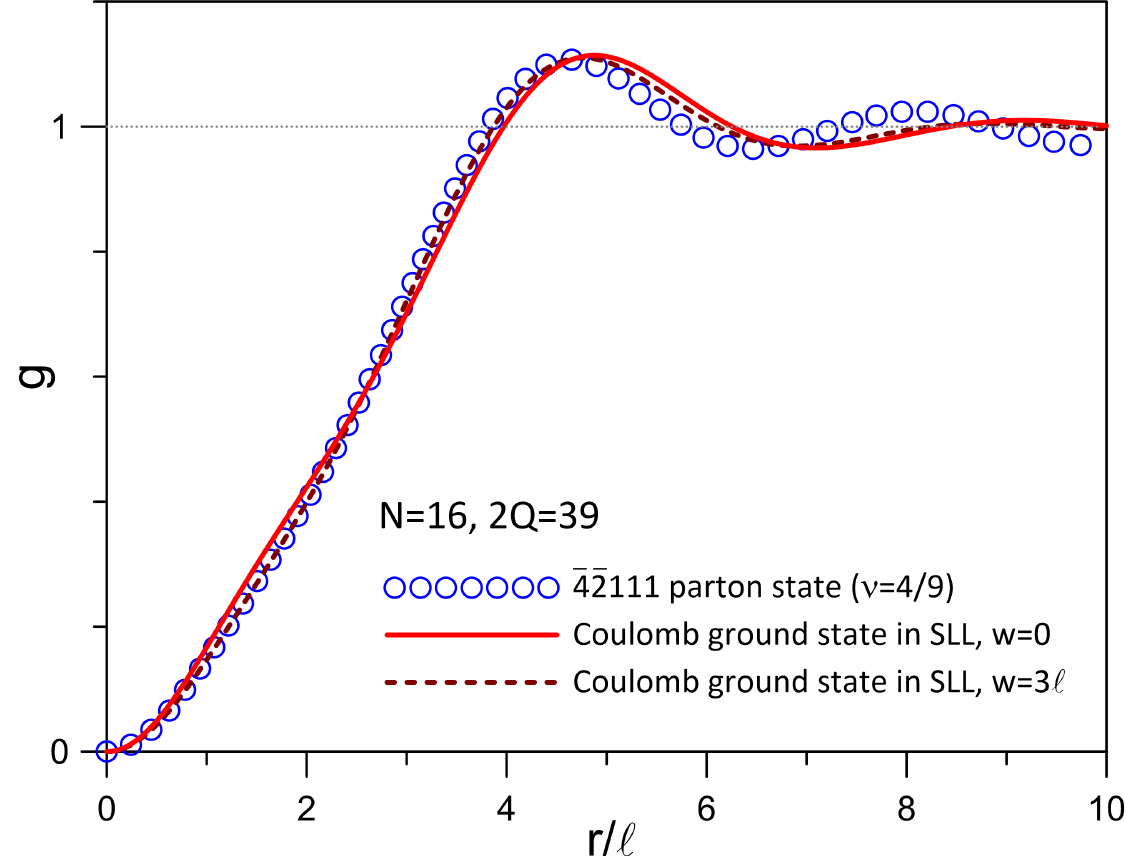} 
\caption{(color online) The pair correlation function $g(r)$ as a function of the arc distance for the exact second Landau level Coulomb ground state obtained using the disk pseudopotentials for width $w=0$ and $w=3\ell$, and the $\bar{4}\bar{2}111$ parton state of Eq.~(\ref{eq:parton_barnbar2111}) for $N=16$ electrons at a flux of $2Q=39$. }
\label{fig:pair_correlations_4_9_finite_w_disk}
\end{center}
\end{figure}

Next, we turn to comparisons of the parton state with the exact SLL Coulomb ground state. The smallest system accessible to exact diagonalization (ED) is that of $N=16$ electrons at a flux of $2Q=39$ which has a Hilbert space dimension of $7\times 10^8$. We have evaluated the ground state for this system with the truncated pseudopotentials from the disk geometry, which differ slightly from the spherical pseudopotentials but are known to provide a more reliable extrapolation to the thermodynamic limit~\cite{Peterson08,Peterson08b}. The exact SLL Coulomb ground obtained by using the truncated disk pseudopotentials has $L=0$. In Fig.~\ref{fig:pair_correlations_4_9_finite_w_disk} we compare the pair-correlation function~\cite{footnote:correlation} of this exact SLL Coulomb ground state with that of the $\bar{4}\bar{2}111$ parton state. Both these pair-correlation functions show oscillations that decay at long distances, which is a typical characteristic of incompressible states~\cite{Kamilla97,Balram15b}. Moreover, the two pair-correlation functions are in reasonable agreement with each other. For completeness, we have also evaluated the exact LLL Coulomb ground state for the same system. The overlaps of the LLL and SLL Coulomb ground states obtained using the disk pseudopotentials is $0.3663$ and their pair-correlation functions are also very different from each other~\cite{SM} which indicate that the nature of the ground state in the two LLs are very different. 

Currently, we do not have a reliable estimate of the thermodynamic values of the gaps predicted by our parton ansatz. However, we can extract the charge and neutral gaps for $N=16$ particles from exact diagonalization of the SLL Coulomb interaction at the parton shift. The charge gap here is defined as $[\mathcal{E}(2Q=40)+\mathcal{E}(2Q=38)-2\mathcal{E}(2Q=39)]/4$, where $\mathcal{E}(2Q)$ is the exact ground state energy at flux $2Q$ and the factor of $4$ in the denominator accounts for the fact that the addition of a single flux quantum in the parton state produces four fundamental quasiholes. The neutral gap is defined as the difference between the two lowest exact energies at the flux of $2Q = 39$. The charge and neutral gaps for $N=16$, evaluated using exact diagonalization with the disk pseudopotentials, are $0.009~e^2/\epsilon \ell$ and $0.005~e^2/\epsilon \ell$ respectively, where $\ell=\sqrt{\hbar c/(eB)}$ is the magnetic length and $\epsilon$ is the dielectric constant of the background host material. 

We next consider the effect of finite width on the system, which we model by taking the transverse wave function to be the ground state for an infinite square quantum well of width $w$ (see SM~\cite{SM} for details). For the disk pseudopotentials, we find that the ground state for $16$ electrons has $L=0$ for (at least) $w\leq5\ell$~\cite{SM}. Moreover, the pair-correlation function of the exact ground state agrees well with that of the parton state for the entire range of widths considered in this work (see Fig.~\ref{fig:pair_correlations_4_9_finite_w_disk} and Ref.~\cite{SM}). Furthermore, we find that the system has robust charge and neutral gaps for all the widths considered. We note that the $L=0$ ground state is delicate. In particular, the exact SLL Coulomb ground state obtained using the spherical pseudopotentials has $L=2$. However, the overlap between the \emph{lowest energy $L=0$ state} obtained using the spherical pseudopotentials and the ground state obtained using the disk pseudopotentials is $0.9692$, which indicates that these two states are close to each other. Encouragingly, with the spherical pseudopotentials, as the quantum well width is increased the ground state turns uniform in the range $w\in(0.5,1]\ell$ and stays uniform for $w\in[1,10]\ell$~\cite{SM}. These results indicate that finite thickness enhances the stability of the parton state.

Now that we have made a case for the plausibility of the $\bar{4}\bar{2}111$ parton state to occur in the SLL, we shall turn to deduce the experimental consequences of this parton ansatz. An additional particle in the factor $\Phi_{\bar{4}}$ has charge $e/9$, whereas that in the factors $\Phi_{\bar{2}}$ and $\Phi_{1}$ has a charge $2e/9$ and $-4e/9$ respectively. All the quasiparticles of the $\bar{4}\bar{2}111$ parton state obey Abelian braid statistics~\cite{Wen91}. The $4/9$ Jain CF state is also an Abelian state and hosts quasiparticles of charge $-e/9$ and $-4e/9$. 

Next, to infer other topological consequences of the $\bar{4}\bar{2}111$ ansatz, we consider the low-energy effective theory of its edge, which is described by the Lagrangian density~\cite{Wen91b,Wen92b,Wen95,Moore98}: 
\be
\mathcal{L} = -\frac{1}{4\pi} K_{\rm IJ}\epsilon^{\alpha\beta\gamma} a_{\alpha}^{\rm I}\partial_{\beta} a_{\gamma}^{\rm J} - \frac{1}{2\pi} \epsilon^{\alpha\beta\gamma}t_{\rm I}A_{\alpha}\partial_{\beta} a_{\gamma}^{\rm I}.
\label{eq_eff_L}
\ee
Here $\epsilon^{\alpha\beta\gamma}$ is the fully anti-symmetric Levi-Civita tensor, $A$ is the vector potential corresponding to the external electromagnetic field, $a$ is the internal gauge field, and we have used the Einstein's convention of summing repeated indices. The integer-valued symmetric $K$ matrix and the charge vector $t$ of Eq.~(\ref{eq_eff_L}) for the parton state are given by (see SM~\cite{SM} for a derivation)
\be
K =   \begin{pmatrix} 
      -2 & -1 &-1 & 0 & 1 \\
      -1 & -2 &-1 & 0 & 1 \\
      -1 & -1 &-2 & 0 & 1 \\
       0 &  0 & 0 &-2 & 1 \\
       1 & 1 &  1 & 1 & 1\\
   \end{pmatrix},\quad
t =    \begin{pmatrix} 
      \,0\, \\
      \,0\, \\
      \,0\, \\
      \,0\, \\
      \,1\, \\
   \end{pmatrix}.   
\ee
The above $K$ matrix has four negative and one positive eigenvalues and thus the $\bar{4}\bar{2}111$ state hosts four upstream and one downstream edge modes.
A naive counting suggests that there are a total of nine edge states for the $\bar{4}\bar{2}111$ ansatz: four from the factor $\Phi_{\bar{4}}$, two from $\Phi_{\bar{2}}$, and one from each factor of $\Phi_1$. However, these edges states are not all independent since the density variations of the five partons must be identified. This results in four constraints and leads to five edge states consistent with that ascertained from the above $K$ matrix.
 
 \emph{Assuming} full equilibration of the edge modes, the thermal Hall conductance $\kappa_{xy}$ at temperatures much smaller than the gap takes a quantized value proportional to the chiral central charge $c_{-}$, which is defined as the difference in the number of downstream and upstream modes: $\kappa_{xy} =c_{-}[\pi^2 k_{\rm B}^2 /(3h)]T$~\cite{Kane97}. For $\bar{4}\bar{2}111$ ansatz, we thus predict a thermal Hall conductance of $\kappa_{xy} =-3[\pi^2 k_{\rm B}^2 /(3h)]T$. The Hall viscosity of the $\bar{4}\bar{2}111$ state is also expected to be quantized~\cite{Read09}: $\eta_{\rm H} = \hbar \rho_{0} \mathcal{S}/4$, where $\rho_{0}=(4/9)/(2\pi \ell^{2})$ is the electron density and $\mathcal{S}=-3$ is the shift of the parton state. The ground-state degeneracy of the parton state on a topologically nontrivial manifold with genus $g$ is $|{\rm Det}(K)|^{g}=18^{g}$. Besides the $n\bar{n}111$ parton states~\cite{Balram19a}, the $\bar{4}\bar{2}111$ ansatz provides another example of a fully spin polarized Abelian FQH state at $\nu=a/b$ (with $a,b$ coprime), which has a ground state degeneracy on the torus that is greater than $b$. 

The $4/9$ Jain CF state is described by the $4\times 4$ $K$ matrix $K=2C_{4}+\mathbb{I}_{4}$, where $C_{k}$ is the $k\times k$ matrix of all ones and $\mathbb{I}_{k}$ is the $k\times k$ identity matrix, and charge vector $t=(1,1,1,1)^{T}$. In contrast to the $\bar{4}\bar{2}111$ state, assuming the absence of edge reconstruction, the $4/9$ Jain CF state has four downstream edge states and no upstream modes. The $4/9$ Jain CF state thus has a thermal Hall conductance of $\kappa_{xy} =4[\pi^2 k_{\rm B}^2 /(3h)]T$. Moreover, the Hall viscosity of the $4/9$ Jain CF state is given by $\eta_{\rm H} = (3/2)\hbar \rho_{0}$, corresponding to shift ${\cal S}=6$. On a manifold of genus $g$, the $4/9$ Jain CF state has a degeneracy of $9^{g}$. 

The presence of upstream neutral modes can be detected in shot noise experiments~\cite{Bid10,Dolev11,Gross12,Inoue14}. Recently, thermal Hall measurements have been carried out at several filling factors in the lowest as well as the second LL~\cite{Banerjee17,Banerjee17b,Srivastav19}. These experiments can be used to test the predictions of the parton theory and therefore can unambiguously distinguish between the topological nature of the 4/9 states in the SLL and the LLL. In particular, including the contributions of the filled LLLs of spin up and spin down, the thermal Hall conductance of the $\bar{4}\bar{2}111$ state in the SLL is $-[\pi^2 k_{\rm B}^2 /(3h)]T$ which is different from what one would expect from the $4/9$ Jain CF state in the SLL, which has $\kappa_{xy} =6[\pi^2 k_{\rm B}^2 /(3h)]T$.

In summary, we have considered the viability of the ``$\bar{4}\bar{2}111$'' parton state for FQHE at $\nu=2+4/9$ where the first signs of incompressibility in the form of minimum in longitudinal resistance have already been observed experimentally~\cite{Choi08,Reichl14}. Interestingly, if FQHE eventually stabilizes at this filling factor, then it is likely to be topologically different from its LLL counterpart at $\nu=4/9$, which is described by a Jain CF state. We also proposed experimental measurements that can reveal the underlying topological structure of the parton state and decisively distinguish it from the $4/9$ state occurring in the lowest Landau level. 

\begin{acknowledgments}
We acknowledge useful discussions with Maissam Barkeshli, Jainendra K. Jain, Mark S. Rudner, Dam T. Son, and Bo Yang. This work was supported by the Polish NCN Grant No. 2014/14/A/ST3/00654 (A. W.). We thank Wroc\l{}aw Centre for Networking and Supercomputing and Academic Computer Centre CYFRONET, both parts of PL-Grid Infrastructure. Some of the numerical calculations reported in this work were carried out on the Nandadevi supercomputer, which is maintained and supported by the Institute of Mathematical Science's High-Performance Computing Center. This research was supported in part by the International Centre for Theoretical Sciences (ICTS) during a visit for the program Novel Phases of Quantum Matter (code: ICTS/topmatter2019/12)
\end{acknowledgments}

\bibliography{biblio_fqhe}

\begin{thebibliography}{78}
\bibitem[{foo()}]{footnote:correlation}
\bibinfo{note}{We compare the pair-correlation functions since, due to
  technical reasons, it has not yet been possible to calculate the overlap of
  the parton state with the state obtained from exact diagonalization for the
  large system size considered in this work.}
\end{thebibliography}
\bibliographystyle{apsrev_nourl}

\clearpage
\begin{widetext}
\begin{center}
\textbf{Supplemental Material for ``Fractional Quantum Hall Effect at $\nu=2+4/9$''}
\end{center}
\end{widetext}

\setcounter{figure}{0}
\setcounter{table}{0}
\setcounter{equation}{0}
\renewcommand\thefigure{S\arabic{figure}}
\renewcommand\thetable{S\arabic{table}}
\renewcommand\theequation{S\arabic{equation}}

In this Supplemental Material (SM) we provide
\begin{itemize}
\item a derivation of the low-energy effective edge theory of the $\bar{4}\bar{2}111$ parton state (Sec.~\ref{sec:eff_edge}),
\item a detailed study of the effect of finite width at $\nu=2+4/9$ (Sec.~\ref{sec:finite_width_4_9_SLL}), and
\item a comparison of the Laughlin, Pfaffian and the $\bar{3}\bar{2}111$ parton states with the exact Coulomb ground state at filling factors $\nu=7/3,~5/2$ and $2+6/13$ respectively (Sec.~\ref{sec:other_states_SLL}).
\end{itemize}

\section{Derivation of the low-energy effective theory of the parton $\bar{4}\bar{2}111$ $\nu=4/9$ edge}
\label{sec:eff_edge}
To derive the low-energy effective theory of the parton $4/9$ edge we closely follow the procedure outlined in the Supplemental Material of Ref.~\cite{Balram18a}. We are interested in the $n=4$ member of the $\bar{n}\bar{2}111$ parton sequence. We shall first discuss the case of general $n$ and then specialize to the case of $n=4$. The unprojected wave function of the $\bar{n}\bar{2}111$ parton state can be re-written as
\begin{equation}
\Psi^{\bar{n}\bar{2}111}_{2n/(5n-2)} = [\Phi_{n}]^{*}[\Phi_{2}]^{*}\Psi^{L}_{1/3},
\label{eq:parton_unprojected_barnbar2111}
\end{equation}
where $\Phi_{n}$ is the Slater determinant state of $n$ filled Landau levels and $\Psi^{L}_{1/3}\equiv \Phi^{3}_{1}$ is the $\nu=1/3$ Laughlin state~\cite{Laughlin83}. This state can be expressed in terms of partons $\wp = f_{1}f_{2}f_{3}$, where the $f_{i}$'s are fermionic partons in the following mean-field states: 
\begin{itemize}
 \item $f_{1}$ is in a $\nu=-n$ integer quantum Hall (IQH) state
 \item $f_{2}$ is in a $\nu=-2$ IQH state
 \item $f_{3}$ is in a $\nu=1/3$ Laughlin state.
\end{itemize}
The charges of these partons are~\cite{Jain89b} $q_{1}=2e/(5n-2)$, $q_{2}=ne/(5n-2)$ and $q_{3}=-6ne/(5n-2)$, where $-e$ is the electron charge. For $n=2$, the parton state of Eq.~(\ref{eq:parton_unprojected_barnbar2111}) describes a non-Abelian state at $\nu=1/2$~\cite{Balram18}. From here on we restrict to $n\neq 2$, wherein the parton state of Eq.~(\ref{eq:parton_unprojected_barnbar2111}) describes an Abelian state with a residual $U(1)\times U(1)$ gauge symmetry associated with the transformations:
\begin{equation}
f_{1}\rightarrow e^{i\theta_{1}} f_{1},~f_{2}\rightarrow e^{-i\theta_{1}+i\theta_{2}} f_{2},~
f_{3}\rightarrow e^{-i\theta_{2}} f_{3}.
\end{equation}
Therefore, we have two internal emergent $U(1)$ gauge fields, denoted by $h_{\mu}$ and $g_{\mu}$, associated with the above transformations. The low-energy effective field theory for this parton mean-field state is described by the Lagrangian density (here and henceforth we have set $e=1$ for convenience)~\cite{Balram18a}:
\begin{eqnarray}
\mathcal{L}&=&\frac{1}{4\pi} \sum_{i=1}^{n}\alpha^{(i)}\partial \alpha^{(i)}+\frac{1}{2\pi}\sum_{i=1}^{n}(h+q_{1}A)\partial \alpha^{(i)}  \nonumber \\
&+& \frac{1}{4\pi} \sum_{j=1}^{2}\beta^{(j)}\partial \beta^{(j)}+\frac{1}{2\pi}\sum_{j=1}^{2}(g-h+q_{2}A)\partial \beta^{(j)}  \nonumber \\
&-& \frac{3}{4\pi} \gamma \partial \gamma+\frac{1}{2\pi}(-g+q_{3}A)\partial \gamma,
\label{eq:Lagrangian_density}
\end{eqnarray}
where $A$ is the external physical electromagnetic vector potential, and $\alpha^{(i)}$, $\beta^{(j)}$ and $\gamma$ are $U(1)$ gauge fields describing the current fluctuations of the IQH and Laughlin states. Furthermore, we have used the short-hand notation $\alpha\partial \alpha\equiv\epsilon^{\mu\nu\lambda}\alpha_{\mu}\partial_{\nu}\alpha_{\lambda}$, where $\epsilon^{\mu\nu\lambda}$ is the fully anti-symmetric Levi-Civita tensor and repeated indices are summed over.  \\

This Chern-Simons theory can be further simplified by integrating out the internal gauge fields $h$ and $g$. Doing so we obtain the following two constraints~\cite{Balram18a}
\begin{equation}
\sum_{i=1}^{n} \alpha^{(i)} = \sum_{j=1}^{2} \beta^{(j)} + c, 
\label{eq:constraint1}
\end{equation}
and
\begin{equation}
\gamma = \sum_{j=1}^{2} \beta^{(j)} + d,
\label{eq:constraint2}
\end{equation}
where $c$ and $d$ are $U(1)$ gauge fields that satisfy $\epsilon^{\mu\nu\lambda} \partial_\nu c_\lambda=0$ and $\epsilon^{\mu\nu\lambda} \partial_\nu d_\lambda=0$. 
Note that when substituting Eq.~(\ref{eq:constraint1}) and (\ref{eq:constraint2}) into Eq.~(\ref{eq:Lagrangian_density}), all terms involving the gauge fields $c$ and $d$ vanish. Thus, we end up with a simplified $U(1)^{n+1}$ Chern-Simons theory which can be described by an integer valued symmetric $(n+1)\times (n+1)$ $K$ matrix~\cite{Balram18a}.

Using the constraints of Eqs.~(\ref{eq:constraint1}) and (\ref{eq:constraint2}) we can eliminate $\alpha^{(n)}$ and $\beta^{(2)}$ by noting that
\begin{equation}
\alpha^{(n)}=\gamma - \alpha^{(1)} - \alpha^{(2)} - \cdots - \alpha^{(n-1)}  +c -d,~
\beta^{(2)} =\gamma - \beta^{(1)} - d.
\end{equation}
Substituting these back into the Lagrangian density given in Eq.~(\ref{eq:Lagrangian_density}) and using the fact that all terms involving the gauge fields $c$ and $d$ vanish, we obtain the following simplified Lagrangian density:
\begin{widetext}
{\small
\begin{eqnarray}
\mathcal{L}&=&\frac{1}{4\pi}\left( \alpha^{(1)}\partial \alpha^{(1)}+\alpha^{(2)}\partial \alpha^{(2)}+\cdots+\alpha^{(n-1)}\partial \alpha^{(n-1)}+
(\gamma - \alpha^{(1)} -\alpha^{(2)}-\cdots-\alpha^{(n-1)}) \partial (\gamma - \alpha^{(1)} -\alpha^{(2)}-\cdots-\alpha^{(n-1)}) \right)  \nonumber \\
&+& \frac{1}{4\pi}\left( \beta^{(1)}\partial \beta^{(1)} + (\gamma-\beta^{(1)})\partial (\gamma-\beta^{(1)})\right) \\
&-& \frac{3}{4\pi} \gamma \partial \gamma + \frac{1}{2\pi}A\partial \gamma.
\label{eq:Lagrangian_density_simplified}
\nonumber 
\end{eqnarray}
}
\end{widetext}
By defining a new set of gauge fields:
\begin{equation}
 (a^{1},a^{2},\cdots,a^{n-1},a^{n},a^{n+1}) = (\alpha^{(1)},\alpha^{(2)},\cdots,\alpha^{(n-1)},\beta^{(1)},\gamma),
\end{equation}
we can write the Lagrangian in the standard form~\cite{Wen91b,Wen92b,Wen95,Moore98}:
\begin{equation}
 \mathcal{L} = -\frac{1}{4\pi} K_{\rm IJ}a^{\rm I}\partial a^{\rm J} + \frac{1}{2\pi} t^{\rm I}A\partial a^{\rm I}.
\end{equation}
Here the charge vector is $t=(0,0,\cdots,0,1)^{\rm T}$ and the $K$ matrix is given by
\begin{equation}
K =   
\begin{pmatrix} 
       -2 & \cdots & -1 & 0& 1\\
       \vdots  & \ddots & \cdots & 0 &1\\
       -1 & \vdots &-2 & 0 & 1 \\
        0 & \cdots & 0 & -2  & 1 \\
        1 & \cdots & 1 & 1 &  1 \\
   \end{pmatrix}. 
\label{eq:Kmatrix_parton_nbarn}   
\end{equation}
The $K$-matrix can be specified succinctly as
\begin{equation}
K_{i,j}=\begin{cases}
-2, & 1\leq i=j \leq n \\
1, & i=n+1 \lor j=n+1 \\
-1, & 1\leq i<j\leq n-1 \\
-1, & 1\leq j<i\leq n-1 \\
0, & {\rm otherwsie} 
\end{cases}
\end{equation}
The filling factor is given by~\cite{Wen95}:
\begin{equation}
\nu =  t^{\rm T}\cdot K^{-1} \cdot t=K^{-1}_{n+1,n+1} = 2n/(5n-2),
\end{equation}
as anticipated from the microscopic wave function. The ground-state degeneracy of the $\bar{n}\bar{2}111$ parton state on a manifold with nontrivial genus $g$ is~\cite{Wen95}:
\begin{equation}
 \text{ground-state degeneracy} = |{\rm Det}(K)|^{g} =(5n-2)^{g}.
\end{equation}
An interesting point to note is that when $n=2m$ is even, then the filling factor $\nu=2n/(5n-2)=2m/(5m-1)$. For these set of fillings, the ground-state degeneracy is $(5n-2)^{g}=2^g (5m-1)^g$. Therefore, for even $n\neq 2$ we get a single component Abelian state at $\nu=a/b$ (with $a,b$ coprime) which has a ground-state degeneracy on the torus ($=2b$) which is greater than $b$ (see Ref.~\cite{Balram19a} for another example of such a state). 

The $K$ matrix of Eq.~(\ref{eq:Kmatrix_parton_nbarn}) has one positive and $n$ negative eigenvalues which indicates that the $\bar{n}\bar{2}111$ state hosts one downstream and $n$ upstream edge modes which gives it a chiral central charge of $1-n$. The particle-hole conjugate of the $\bar{n}\bar{2}111$ state, which occurs at filling factor $1-2n/(5n-2)=(3n-2)/(5n-2)$, can be viewed as a \emph{fully chiral} state with $n$ downstream modes which results in a chiral central charge of $n$. The charges and the braiding statistics of the quasiparticles of the $\bar{n}\bar{2}111$ state can be ascertained from its $K$ matrix following the work of Ref.~\cite{Wen95}. 

\subsection{Coupling to curvature and shift~\cite{Wen92}}
To compute the shift $\mathcal{S}$~\cite{Wen92} of the $\bar{n}\bar{2}111$ state on the sphere from its topological field theory, we need to couple it to the curvature. For a state filling the $n^{\rm th}$ Landau level and described by the $U(1)$ gauge field $\Gamma$, upon including the coupling to curvature, the effective Lagrangian density is modified by the addition of the following term~\cite{Wen92}:
\begin{equation}
\delta\mathcal{L} = \frac{1}{2\pi} s \omega \partial \Gamma,
\end{equation}
where $\omega$ is the spin connection and the spin is $s=(n-1/2)$. Including the coupling to curvature for the $\bar{n}\bar{2}111$ parton state amounts to adding the following additional terms in the Lagrangian density of Eq.~(\ref{eq:Lagrangian_density}):
\begin{equation}
\delta\mathcal{L} = -\frac{1}{2\pi} \sum_{i=1}^{n} (i-1/2)\omega \partial \alpha^{(i)} 
-\frac{1}{2\pi} \sum_{j=1}^{2} (j-1/2)\omega \partial \beta^{(j)} 
+ \frac{1}{2\pi} \frac{3}{2} \omega \partial \gamma. 
\end{equation}
Using the constraints of Eqs.~(\ref{eq:constraint1}) and (\ref{eq:constraint2}) we again eliminate $\alpha^{(n)}$ and $\beta^{(2)}$ to end up with the following additional term in the Lagrangian density of Eq.~(\ref{eq:Lagrangian_density_simplified}) which describes coupling of the $\bar{n}\bar{2}111$ parton state to the curvature:
\begin{widetext}
\begin{eqnarray}
\delta\mathcal{L} &=& -\frac{1}{2\pi} \left( (1/2)\omega \partial \alpha^{(1)} + (3/2)\omega \partial \alpha^{(2)}+ \cdots + ((2n-3)/2)\omega \partial \alpha^{(n-2)}+
((2n-1)/2) \omega \partial (\gamma - \alpha^{(1)} -\alpha^{(2)}-\cdots-\alpha^{(n-1)})\right) \nonumber \\
&-&\frac{1}{2\pi} \left( (1/2)\omega \partial \beta^{(1)} + (3/2)\omega \partial (\gamma - \beta^{(1)})\right)+ \frac{1}{2\pi} \frac{3}{2} \omega \partial \gamma \nonumber \\
&=& \frac{1}{2\pi} \left( (n-1)\omega \partial \alpha^{(1)} + (n-2)\omega \partial \alpha^{(2)} +\cdots + \omega \partial \alpha^{(n-1)}+ \omega \partial \beta^{(1)}-((2n-1)/2)\omega \partial \gamma \right) =\frac{1}{2\pi} \mathfrak{s}^{\rm I}\omega\partial a^{\rm I}.
\end{eqnarray}
\end{widetext}
Here we have defined the spin vector $\mathfrak{s}=((n-1),(n-2),\cdots,1,1,-((2n-1)/2))^{\rm T}$. The shift $\mathcal{S}$ on the sphere is given by~\cite{Wen95}:
\begin{equation}
 \mathcal{S}=\frac{2}{\nu} t^{\rm T}\cdot K^{-1} \cdot \mathfrak{s}=1-n,
\end{equation}
which is consistent with the value ascertained from the microscopic wave function.

\subsection{Specializing to the $n=4$ case for the $\nu=4/9$ state}
For the case of $n=4$ the charge vector is $t=(0,0,0,0,1)^{\rm T}$ and the $K$-matrix following Eq.~(\ref{eq:Kmatrix_parton_nbarn}) is given by
\begin{equation}
K  =
\begin{pmatrix} 
      -2 & -1 &-1 & 0 & 1 \\
      -1 & -2 &-1 & 0 & 1 \\
      -1 & -1 &-2 & 0 & 1 \\
       0 &  0 & 0 &-2 & 1 \\
       1 & 1 &  1 & 1 & 1\\
   \end{pmatrix}. 
\end{equation} 
The filling fraction is:
\begin{equation}
\nu =  t^{\rm T}\cdot K^{-1} \cdot t=K^{-1}_{5,5} = 4/9,
\end{equation}
as expected. The ground state-degeneracy on a manifold with genus $g$ is:
\begin{equation}
 \text{ground-state degeneracy} = |{\rm Det}(K)|^{g} =18^{g}.
\end{equation}
This $K$ matrix has one positive and four negative eigenvalues which indicates that the $\bar{4}\bar{2}111$ state hosts one downstream and four upstream edge modes. The charges of the quasiparticles and their braiding statistics can be read off from the $K$ matrix~\cite{Wen95}.

The examples of the $\bar{4}\bar{2}111$ state considered in the present work and the $n\bar{n}111$ states considered in Ref.~\cite{Balram19a} show that in general a single-component Abelian fractional quantum Hall state at filling factor $p/q$ (with $p,q$ coprime) has the following properties:
\begin{itemize}
\item Its ground state degeneracy on the torus is $kq$, where $k$ is a natural number. 
\item The minimally charged quasiparticle has a charge $(-e)/(mq)$, where $m$ is a natural number, and $(-e)$ is the charge of the electron.
\end{itemize}
In general, $m\neq k$. The $\bar{4}\bar{2}111$ state, which occurs at $\nu=4/9$, has $k=2$, and $m=1$. The $n\bar{n}111$ states, which occur at $\nu=1/3$ for all $n$, have $k=n^2$ and $m=n$~\cite{Balram19a}. Although these examples suggest that $k$ is a multiple of $m$, we do not know if there exists a relation between $k$ and $m$ for a generic single-component Abelian state.

\section{Effect of finite width at $\nu=2+4/9$}
\label{sec:finite_width_4_9_SLL}
We model the effect of finite thickness on the system by considering an infinite square well of width $w$. The disk pseudopotentials of the Coulomb interaction, which are modified by the finite width $w$, in the Landau level indexed by $n$ are given by 
\begin{eqnarray}
V_{m} &=& \int_{0}^{\infty} qdq~V(q)\left[L_{n}\left(\frac{q^{2}\ell^{2}}{2}\right)\right]^{2}L_{m}(q^{2}\ell^{2}) e^{-q^{2}\ell^{2}}, \nonumber \\
V(q)&=&\frac{e^{2}}{\epsilon q}\int^{w/2}_{-w/2} dz_{1}\int^{w/2}_{-w/2} dz_{2}~|\xi(z_{1})|^{2}|\xi(z_{1})|^{2}e^{-q|z_{1}-z_{2}|} \nonumber, \\
\xi(z)&=&\sqrt{\frac{2}{w}}\cos\left(\frac{\pi z}{w} \right),
\end{eqnarray}
where $\xi(z)$ is the wave function in the transverse direction and $L_{n}(x)$ is the $n$th order Laguerre polynomial. Analogously, one can also calculate the spherical pseudopotentials for a system with finite width. We shall present results for $N=16$ electrons at the $\bar{4}\bar{2}111$ flux $2Q=39$ obtained on the spherical geometry using both the disk and spherical pseudopotentials.

\begin{figure}[htpb!]
\begin{center}
\includegraphics[width=0.47\textwidth,height=0.29\textwidth]{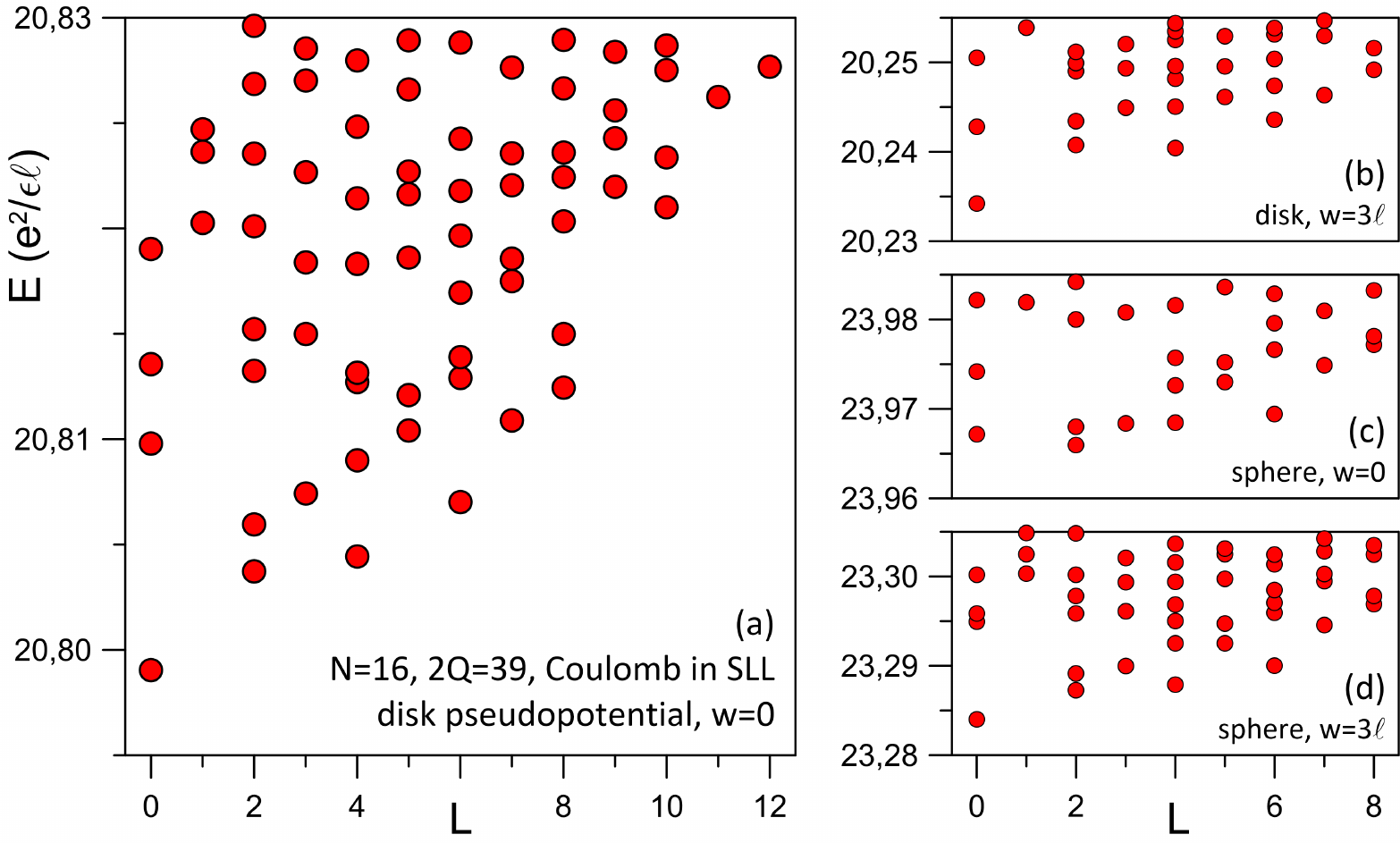} 
\caption{(color online) Exact second Landau level Coulomb spectra $N=16$ electrons at a flux of $2Q=39$ in the spherical geometry evaluated using the a) disk pseudopotentials, b) disk pseudopotentials for width $w=3\ell$, c) spherical pseudopotentials and d) spherical pseudopotentials for width $w=3\ell$. }
\label{fig:1639_spectrum_width_LL1_sphere_disk}
\end{center}
\end{figure}
The exact second Landau level Coulomb spectra obtained using the disk and spherical pseudopotentials for $w=0$ and $w=3\ell$ for $N=16$ electrons at a flux $2Q=39$ is shown in Fig.~\ref{fig:1639_spectrum_width_LL1_sphere_disk}. The ground state, as well as the structure of the low-energy spectra obtained using the disk and spherical pseudopotentials, are quite different from each other, which indicates that finite-size effects are quite strong in the second Landau level. However, we find that the overlap between the \emph{lowest energy $L=0$} state obtained using the spherical pseudopotentials and the ground state obtained using the disk pseudopotentials is $\geq0.97$ for $w\in[0,5]\ell$. This shows that while the choice of pseudopotentials may affect selecting the global ground state in the SLL, the lowest energy $L=0$ state itself is insensitive to the choice. 

\begin{figure}[htpb!]
\begin{center}
\includegraphics[width=0.47\textwidth,height=0.31\textwidth]{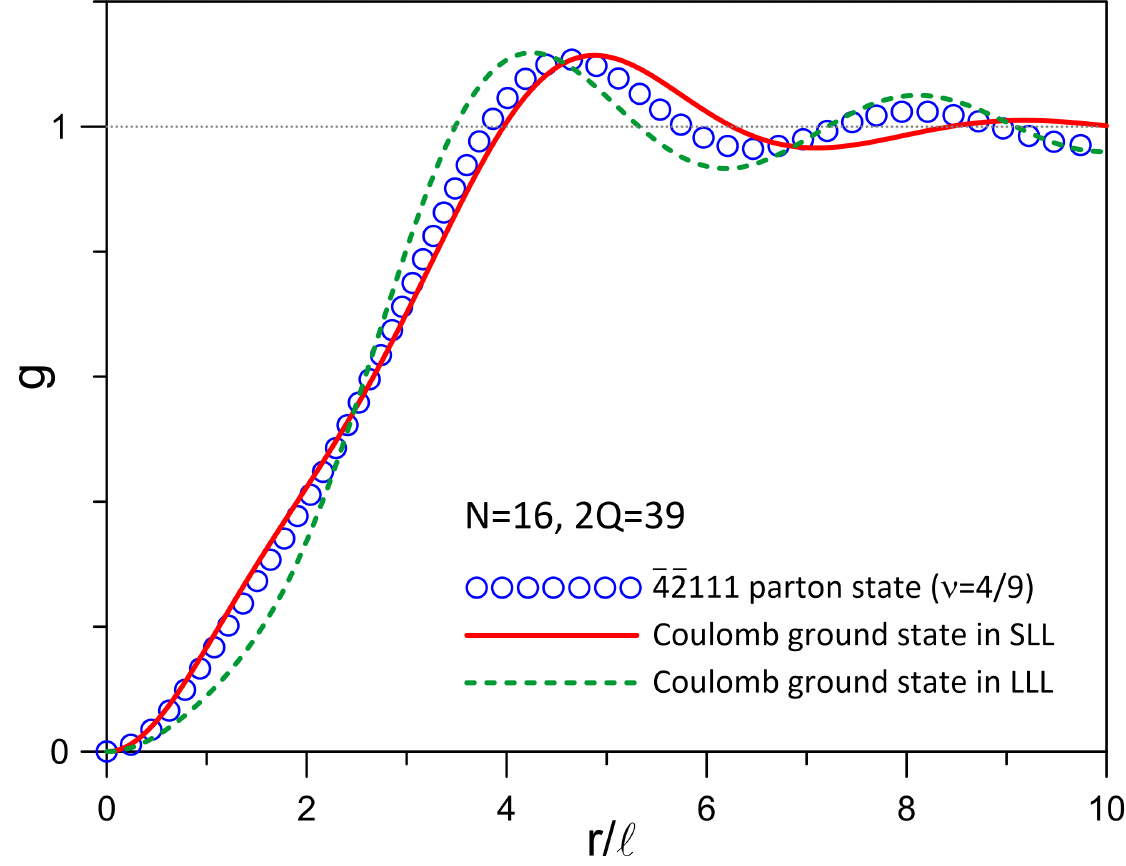} 
\caption{(color online) The pair correlation function $g(r)$ as a function of the arc distance for the exact second Landau level Coulomb ground state obtained using the disk pseudopotentials, and the $\bar{4}\bar{2}111$ state for $N=16$ electrons at a flux of $2Q=39$. For comparison we also show $g(r)$ for the exact LLL Coulomb ground state for the same system.}
\label{fig:pair_correlations_4_9}
\end{center}
\end{figure}

Using the disk pseudopotentials, we find that the ground state is uniform for $w\in [0,5\ell]$. Moreover, the pair-correlation function $g(r)$ of the Coulomb ground state in the SLL for different widths have a visible but small difference indicating that they are fairly close to one another. Furthermore, the pair correlation function of the exact ground state obtained using the disk pseudopotentials agrees well with that of the $\bar{4}\bar{2}111$ parton state for the entire range of widths considered. With increasing width, the Coulomb $g(r)$ looks more similar to the parton $g(r)$ indicating that finite thickness enhances the stability of the parton state (see Fig. 2 of the main text). With the spherical pseudopotentials, the ground state has $L=2$ for $w=0$ and $w=0.5\ell$. Encouragingly, as the quantum well width is increased, the ground state crosses from $L=2$ to $L=0$ in the range $w\in (0.5,1]\ell$ and stays uniform for $w\in [1,10]\ell$. These results indicate that finite thickness aids in stabilizing the $\bar{4}\bar{2}111$ parton state in the second Landau level. For comparison in Fig.~\ref{fig:pair_correlations_4_9}) we also show $g(r)$ for the exact LLL Coulomb ground state which indicates that the ground state in the lowest two LLs for this system are very different from each other.

We have also evaluated the charge gap as a function of finite width using both the disk and spherical pseudopotentials. To obtain the charge gap, we calculate the average charging energy $\mathcal{C}$ of the full LL per pair:
\begin{equation}
\mathcal{C} = \sum_{-l\leq i,j \leq l} \frac{\langle i,j| V | j,i \rangle}{(2l+1)^2},
\label{eq:charging_energy}
\end{equation}
where $\langle i,j| V | j,i \rangle$ is the direct matrix element of the interaction $V$, $l=Q+n$ is the shell angular momentum and $n$ is the LL index. Using the spherical symmetry of a filled LL, $\mathcal{C}$ can be simplified to
\begin{equation}
\mathcal{C} = \sum_{-l\leq j \leq l} \frac{\langle i,j| V | j,i \rangle}{(2l+1)},
\label{eq:charging_energy_simplified_spherical_symmetric}
\end{equation}
where $i$ can be chosen to be any of the orbitals, e.g., the one with $m=-l$ or $m=l$. The charging energies associated with a model pseudopotential Hamiltonian with a single nonzero pseudopotential $H=\delta_{m,j}$ can be expressed analytically as:
\begin{equation}
\mathcal{C}_{j} = \frac{(4l-2j+1)}{(2l+1)^{2}}.
\label{eq:charging_energy_single_Vm}
\end{equation}
Note that the $\mathcal{C}_{j}$’s are normalized, i.e., $\sum_{j=0}^{2l} \mathcal{C}_{j} = 1$. The charging energy for an arbitrary interaction specified by the set of pseudopotentials $\{V_{m}\}$ is therefore:
\begin{equation}
\mathcal{C} = \sum_{m=0}^{2l}V_{m} \frac{(4l-2m+1)}{(2l+1)^{2}}.
\label{eq:charging_energy_arbitrary_interaction}
\end{equation}

For the spherical pseudopotentials the average charging energy for the $1/r$ Coulomb interaction is equal to the inverse radius, i.e.,
\begin{equation}
\mathcal{C}^{\rm sphere}\left(\frac{1}{r}\right) = \frac{e^{2}}{\epsilon R} = \frac{e^{2}}{\epsilon \ell\sqrt{Q}} = \frac{e^{2}}{\epsilon\ell \sqrt{l-n}}.
\label{eq:charging_energy_Coulomb_interaction}
\end{equation}
In particular, the charging energy for our system of interest at $2l=39$ for $w=0$ is $e^{2}/(\epsilon \ell \sqrt{39/2})$ and $e^{2}/(\epsilon \ell \sqrt{37/2})$ in the LLL and SLL respectively. In particular, the charging energy for our system of interest at $2l=39$ for $w=0$ is $e^{2}/(\epsilon \ell \sqrt{39/2})$ and $e^{2}/(\epsilon \ell \sqrt{37/2})$ in the LLL and SLL respectively. Note that for the Coulomb interaction $V(r)=1/r$, Eq.~(\ref{eq:charging_energy_Coulomb_interaction}) leads to a sum rule for the Coulomb pseudopotentials $\{V^{\rm Coulomb}_{m}\}$'s, namely $\sum_{m=0}^{2l} V^{\rm Coulomb}_{m}(4l-2m+1) = (2l+1)^2/R$. For the planar disk pseudopotentials the average charging energy for the $1/r$ Coulomb interaction in the lowest two LLs, in terms of the shell angular momentum $l$, is given by
\begin{eqnarray}
\mathcal{C}^{\rm disk,~LLL}\left(\frac{1}{r}\right) &=& \frac{(3+4(2l))\Gamma[2l+3/2]}{3(2l+1)^{2}\Gamma[2l+1]}\frac{e^{2}}{\epsilon \ell} \\
\mathcal{C}^{\rm disk,~SLL}\left(\frac{1}{r}\right) &=& \frac{(256(2l)^3+20(2l)-33)\Gamma[2l-1/2]}{192(2l+1)^{2}\Gamma[2l+1]}\frac{e^{2}}{\epsilon \ell}  \nonumber,
\label{eq:charging_energy_Coulomb_interaction_disk}
\end{eqnarray}
where $\Gamma[x]$ is the Gamma function.

The value of the charging energy of the finite-width interaction has to be numerically evaluated since no closed-form analytic expression for the finite-width pseudopotentials is known. We find that the $\mathcal{C}^{\rm disk}(1/r)<\mathcal{C}^{\rm sphere}(1/r)$ since the disk pseudopotentials decrease more quickly with relative angular momentum $m$ than the spherical pseudopotentials. Throughout this work we consider only fully spin-polarized electrons, therefore, the electron-electron interaction energies only depend on the odd pseudopotentials. However, the average charging energy, which depends on the direct matrix element of the interaction, is a function of both the even and odd pseudopotentials. Note that it is customary to quote the value of $2l$ itself as the flux $2Q$ since the numerics on the SLL are simulated using effective pseudopotentials in the LLL (and in the LLL $2l=2Q$). 

Using the average charging energies we compute the charged gaps as follows:
\begin{eqnarray}
\mathcal{E}(2Q-1)&=&E(2Q-1)-(N^{2}-(n_{q} e_{q})^2) \frac{\mathcal{C}(2Q-1)}{2}, \nonumber \\ 
\mathcal{E}(2Q   )&=&E(2Q)-(N^{2}           ) \frac{\mathcal{C}(2Q)}{2},  \nonumber \\ 
\mathcal{E}(2Q+1)&=&E(2Q+1)-(N^{2}-(n_{q}e_{q})^2)\frac{\mathcal{C}(2Q+1)}{2},  \nonumber \\ 
\Delta^{\rm charge}&=&\frac{\mathcal{E}(2Q-1)+\mathcal{E}(2Q+1)-2\mathcal{E}(2Q)}{n_{q}},
\label{eq:charge_gap}
\end{eqnarray}
where $E(2Q)$ are ground state energies obtained from exact diagonalization of $N$ electrons at flux $2Q$, $n_{q}=4$ is the number of quasiholes or quasiparticles produced per flux quantum and $e_{q}=1/9$ is the magnitude charge of the quasihole or quasielectron in units of the electronic charge. The $N^2$ term includes the background contribution, and the $(n_{q} e_{q})^2$ term accounts for the fact that the background is different when some charge is accumulated in form of quasiholes or quasielectrons. For the spherical pseudopotentials for $w=0$ and $w=0.5\ell$ we take the lowest energy $L=0$ state as the ground state to evaluate the charge gap. The neutral gap is defined as the difference between the two lowest energy states at the flux $2Q$ corresponding to the ground state. We multiply the energies by a factor of $\sqrt{(2l-2n)\nu/N}$ that corrects for the deviation of the electron density of a finite system from its $N\rightarrow \infty$ value before extrapolating them to the thermodynamic limit~\cite{Morf86}.  

\begin{figure}[htpb!]
\begin{center}
\includegraphics[width=0.47\textwidth,height=0.31\textwidth]{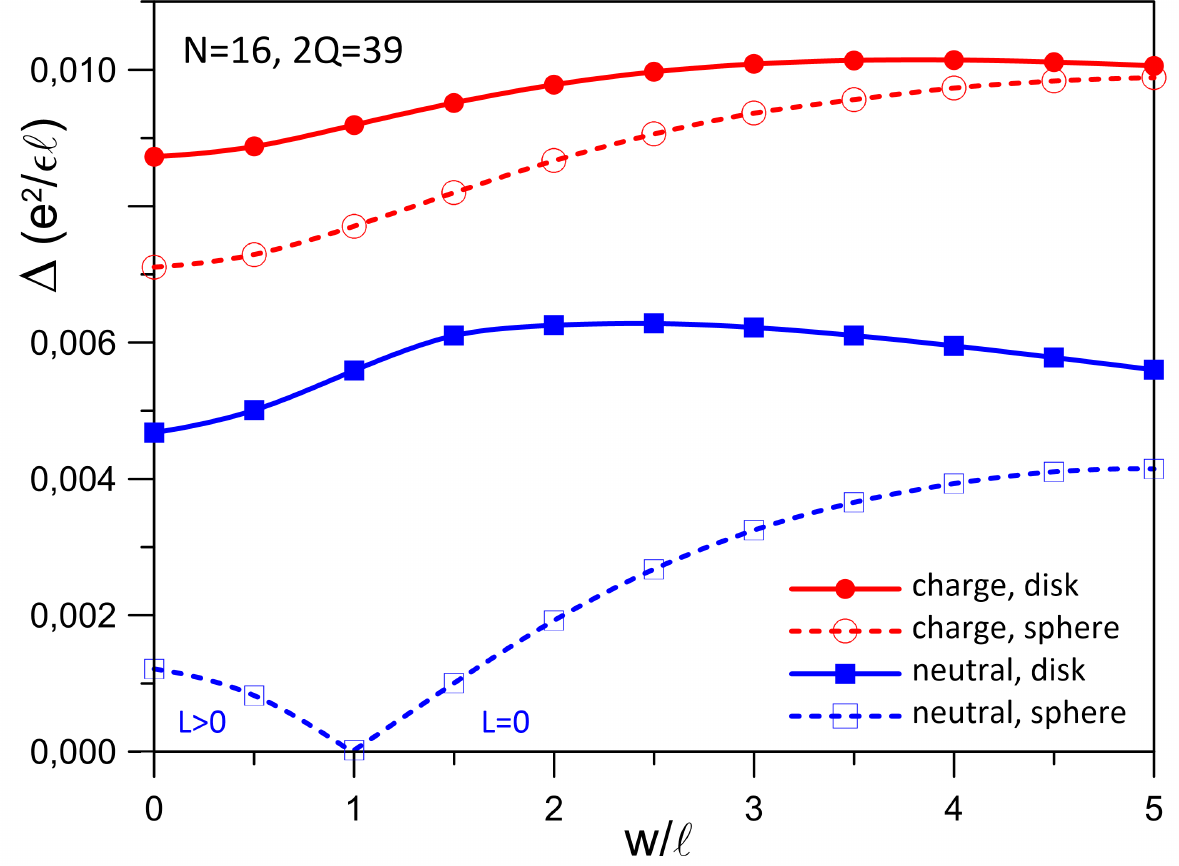} 
\caption{(color online) Charge and neutral gaps for $N=16$ electrons at a flux of $2Q=39$ evaluated in the spherical geometry using the disk and spherical pseudopotentials for various widths $w$. }
\label{fig:1639_charge_neutral_gaps_sphere_disk}
\end{center}
\end{figure}

We find that both the charge and neutral gaps evaluated using the disk pseudopotentials first increase as the well-width is increased and then slowly decreases as the well-width is further increased. In the entire range of widths we considered, the charge and neutral gaps obtained from the disk pseudopotentials are of the order of  $0.01~e^{2}/(\epsilon\ell)$ and $0.005~e^{2}/(\epsilon\ell)$ respectively (see Fig.~\ref{fig:1639_charge_neutral_gaps_sphere_disk}). With the spherical pseudopotentials, the charge gap evaluated using the lowest energy $L=0$ state is negative for $w=0$ and $w=0.5\ell$ but in Fig.~\ref{fig:1639_charge_neutral_gaps_sphere_disk} we have shown only its magnitude to follow the conventional way of indicating phase transitions. In the range from $w=1\ell$ to $w=5\ell$, we find that both the charge and neutral gaps evaluated using the spherical pseudopotentials also monotonically increase as the well-width is increased These results corroborate the fact that the finite thickness of the quantum well enhances the stability of the $2+4/9$ FQHE.

\begin{figure}[htpb!]
\begin{center}
\includegraphics[width=0.47\textwidth,height=0.41\textwidth]{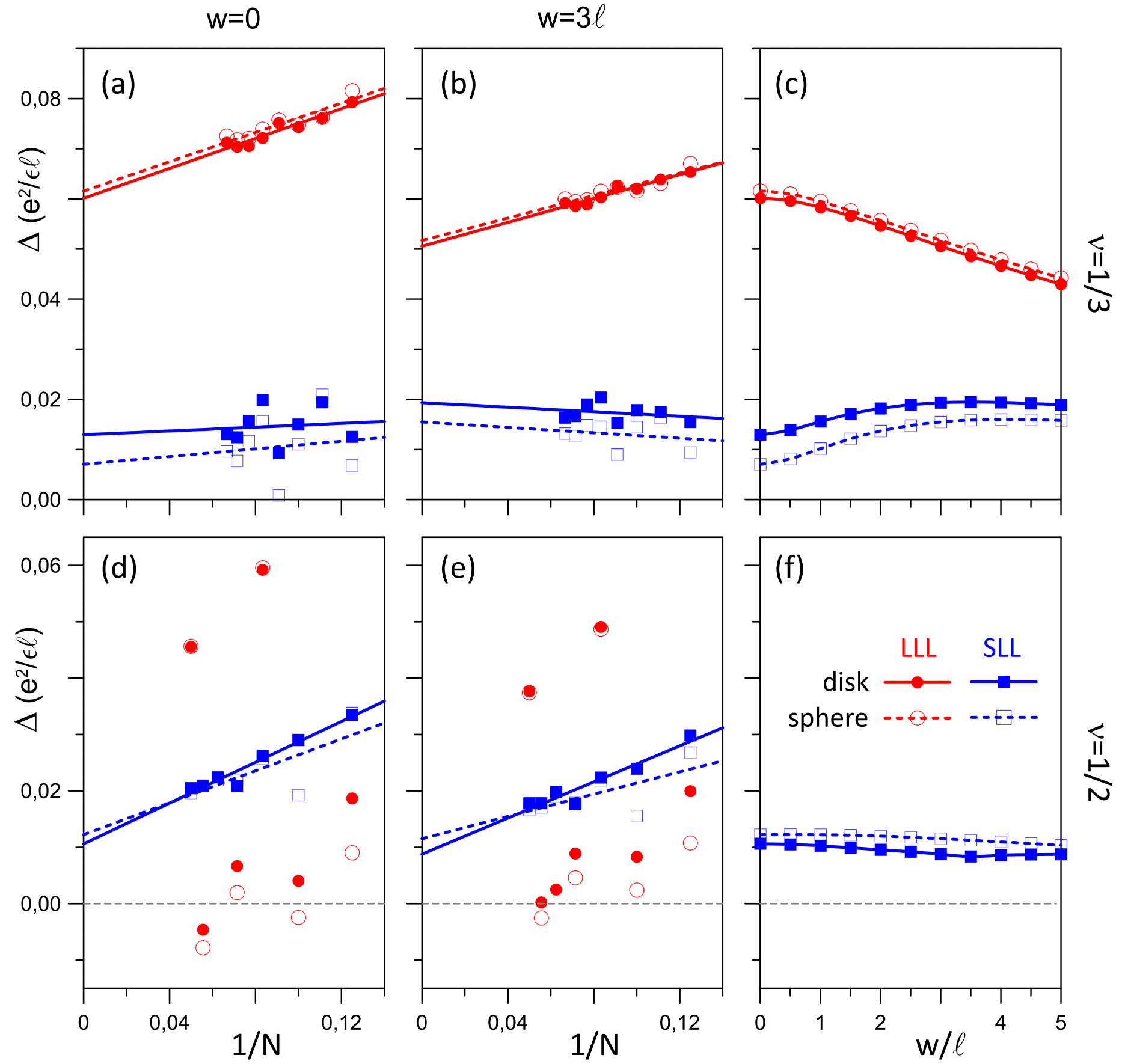} 
\caption{(color online) Neutral gaps in the second Landau level (blue) evaluated in the spherical geometry using the disk and spherical pseudopotentials for various widths $w$ at the $1/3$ Laughlin (top panels) and $1/2$ Pfaffian (bottom panels) fluxes. Panels a) and d) [b) and e)] show linear extrapolation of the neutral gap as a function of $1/N$ for $w=0$ [$w=3\ell$]. The extrapolated gaps are shown in panels c) and f). For completeness, we have also shown the corresponding lowest Landau level gaps (red).} 
\label{fig:Laughlin_Pfaffian_neutral_gaps_sphere_disk}
\end{center}
\end{figure}

\begin{figure}[htpb!]
\begin{center}
\includegraphics[width=0.47\textwidth,height=0.41\textwidth]{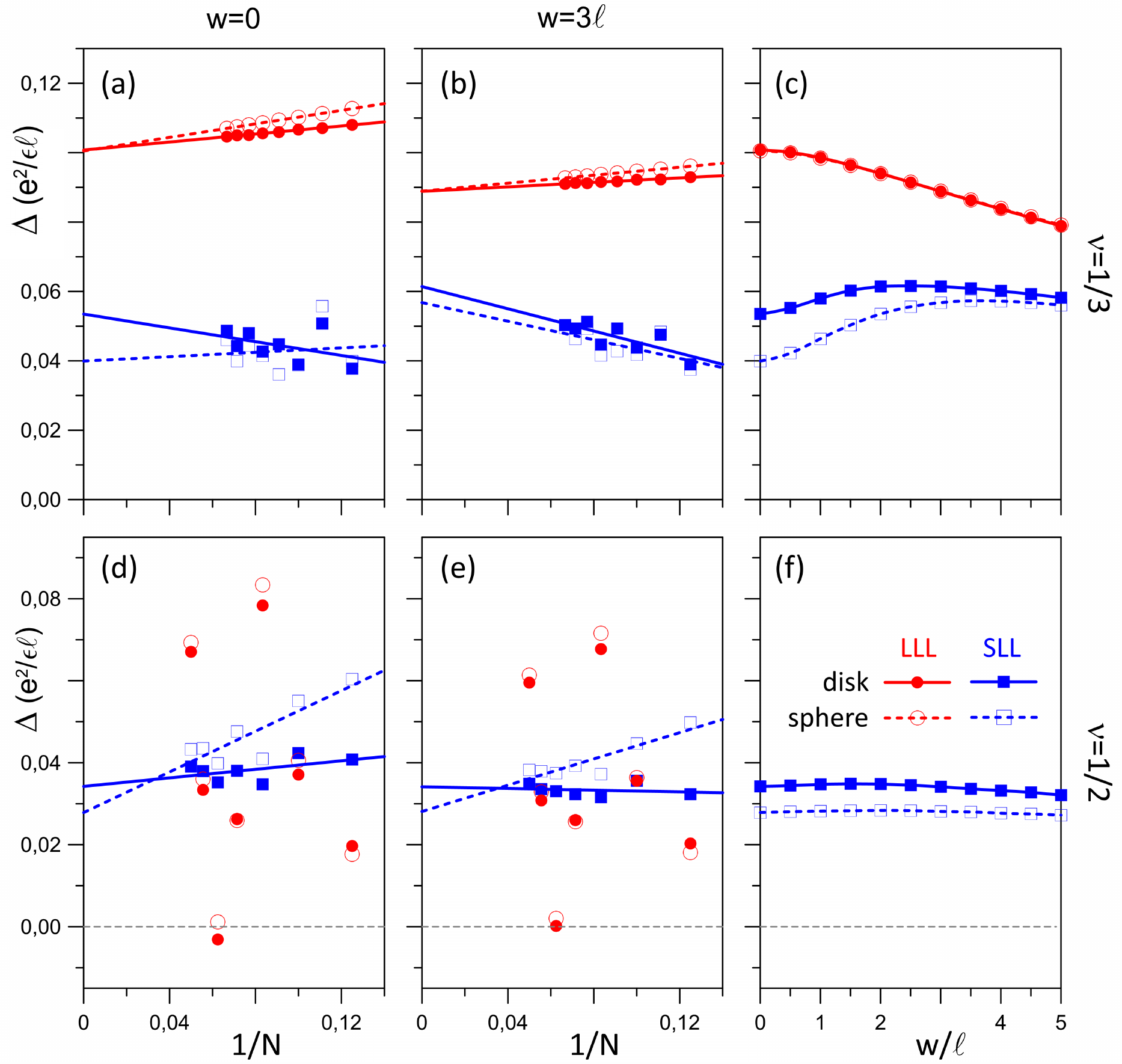} 
\caption{(color online) Charge gaps in the second Landau level (blue) evaluated in the spherical geometry using the disk and spherical pseudopotentials for various widths $w$ at the $1/3$ Laughlin (top panels) and $1/2$ Pfaffian (bottom panels) fluxes. Upon the insertion of a single flux quantum, the $1/3$ Laughlin state produces a single quasihole while the $1/2$ Pfaffian state produces two quasiholes. Thus we use $n_{q}=1$ and $n_{q}=2$ for the $1/3$ Laughlin and $1/2$ Pfaffian states respectively [see Eq.~(\ref{eq:charge_gap})]. Panels a) and d) [b) and e)] show linear extrapolation of the charge gap as a function of $1/N$ for $w=0$ [$w=3\ell$]. The extrapolated gaps are shown in panels c) and f). For completeness, we have also shown the corresponding lowest Landau level gaps (red).} 
\label{fig:Laughlin_Pfaffian_charge_gaps_sphere_disk}
\end{center}
\end{figure}

For comparison, in Figs.~\ref{fig:Laughlin_Pfaffian_neutral_gaps_sphere_disk} and \ref{fig:Laughlin_Pfaffian_charge_gaps_sphere_disk} we show the neutral and charge gaps obtained using the disk and spherical pseudopotentials for various widths at the Laughlin (extrapolated from $N=8,9,\cdots,15$) and Pfaffian (extrapolated from $N=8,10,\cdots,20$) fluxes in the SLL (see also Refs.~\cite{Morf98, Wojs09}). For completeness, on the same plots, we have also included the corresponding LLL gaps. The ground state at the half-filled LLL is a compressible Fermi liquid state of composite fermions. Therefore, the gap in the LLL at the Pfaffian flux is irregular as a function of $1/N$ (the gap is only large when the flux and number of electrons are aliased with a Jain CF state), so we have not extrapolated it to the thermodynamic limit. We find that the gaps for the largest systems at $7/3$ and $5/2$ are a factor of 2-4 times larger than those for the $16$-particle $2+4/9$ state. This indicates that the $2+4/9$ state is more fragile as compared to the experimentally observed $7/3$ and $5/2$ states.

\section{Comparison between some candidate and exact ground states in the second Landau level}
\label{sec:other_states_SLL}
The wave function for the $\bar{4}\bar{2}111$ parton state is most easily evaluated in the LLL in the first-quantized coordinate space (as opposed to the Hilbert space of Fock states). Therefore, to evaluate its second LL Coulomb energy, we use the effective interaction of Ref.~\cite{Shi08} to simulate the physics of the second LL in the LLL. This effective interaction has nearly the same Haldane pseudopotentials as that of the Coulomb interaction in the second LL. Including the contribution of the positively charged background, the per-particle density corrected~\cite{Morf86} energy of the $\bar{4}\bar{2}111$ parton state for the effective interaction for the system of $N=16$ particles is $-0.3839(1)$ (the number in the parenthesis is the Monte Carlo uncertainty in the energy estimate) while the exact energy of the lowest-lying $L=0$ state is $-0.3858$, both in Coulomb units of $e^{2}/(\epsilon\ell)$. The level of agreement (within $0.49\%$) between these two numbers is comparable with that of other trial states in the second LL~\cite{Hutasoit16, Balram18a}. The rest of the section is devoted to showing results for other candidate states and their comparison with the exact second LL Coulomb ground states.

\begin{figure}[htpb!]
\begin{center}
\includegraphics[width=0.47\textwidth,height=0.31\textwidth]{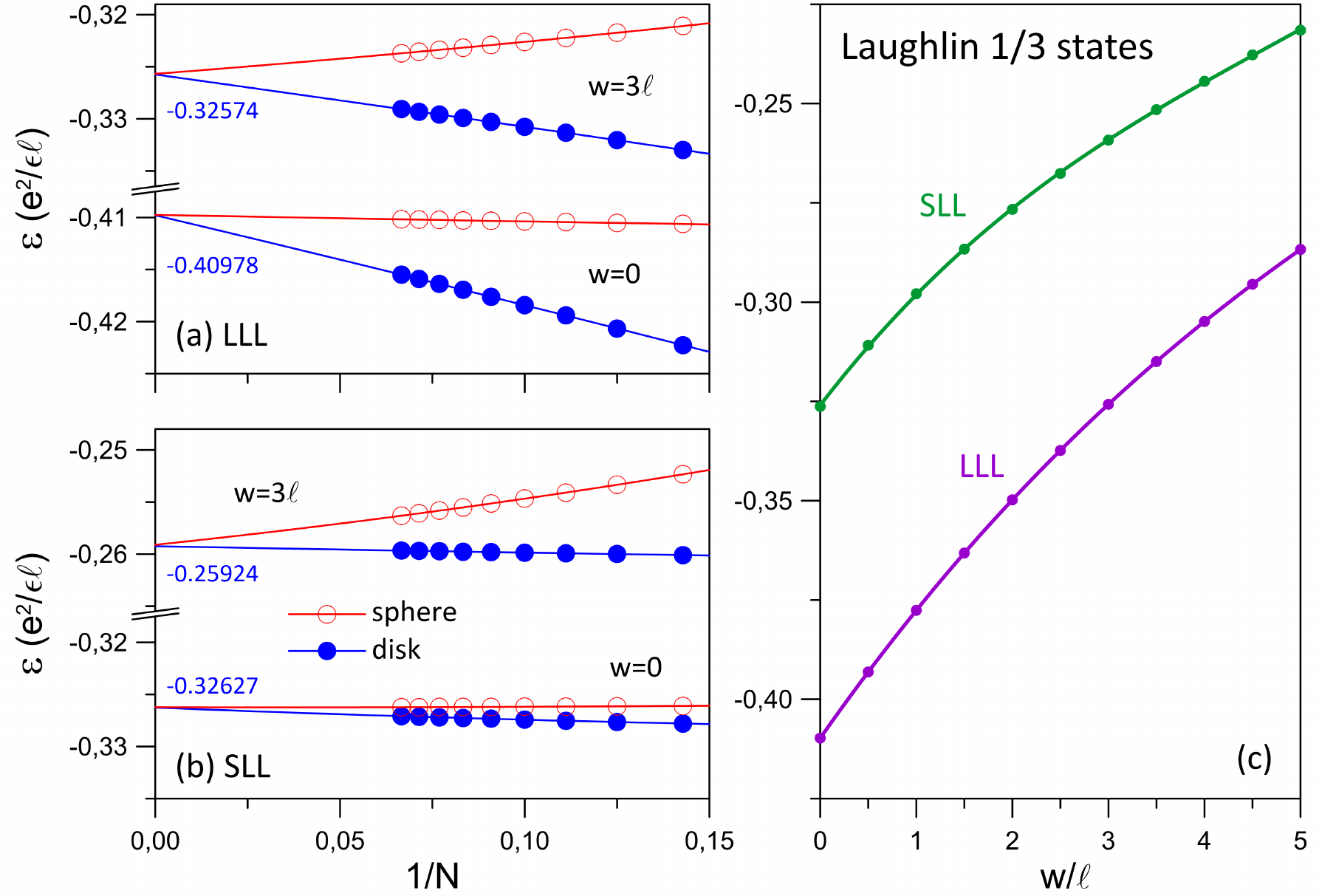} 
\caption{(color online) Thermodynamic extrapolations of the Coulomb per-particle energies for the $\nu=1/3$ Laughlin state in the lowest and second Landau level obtained using the spherical and disk pseudopotentials at various finite widths. The extrapolated energies are shown in panel c) are obtained for the spherical pseudopotentials from quadratic fits in $1/N$. }
\label{fig:extrapolations_energies_1_3_Laughlin}
\end{center}
\end{figure}

\begin{figure}[htpb!]
\begin{center}
\includegraphics[width=0.47\textwidth,height=0.31\textwidth]{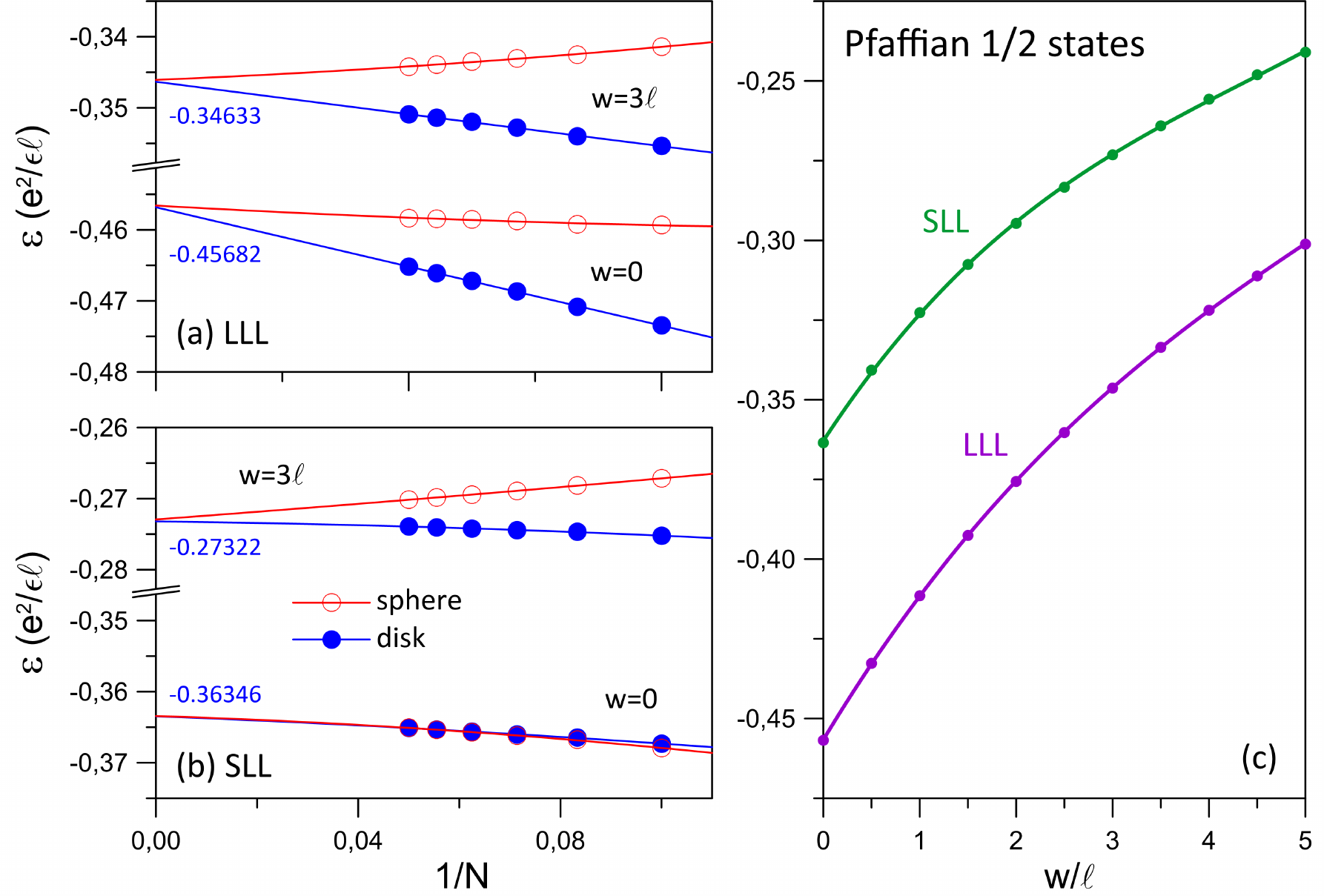} 
\caption{(color online) Thermodynamic extrapolations of the Coulomb per-particle energies for the $\nu=1/2$ Pfaffian state in the lowest and second Landau level obtained using the spherical and disk pseudopotentials at various finite widths. The extrapolated energies shown in panel c) are obtained for the spherical pseudopotentials from quadratic fits in $1/N$.}
\label{fig:extrapolations_energies_1_3_Pfaffian}
\end{center}
\end{figure}

In Tables~\ref{energies_overlaps_SLL_Laughlin} and \ref{energies_overlaps_SLL_MR_Pfaffian} we compare the overlap and energies of the Coulomb ground state in the two lowest Landau levels with the Laughlin and Pfaffian state respectively. For the largest system of $N=15$ electrons considered in this work, the energy of the Laughlin state differs from the exact second LL Coulomb energy by about $1.1\%$. Similarly, the energy of the Pfaffian state of $N=20$ electrons is within $0.6\%$ of the exact second LL Coulomb ground state. These numbers are of the same order of magnitude as those of the $\bar{4}\bar{2}111$ parton state shown above. For completeness, in Figs.~\ref{fig:extrapolations_energies_1_3_Laughlin} and \ref{fig:extrapolations_energies_1_3_Pfaffian} we show the thermodynamic extrapolation and the extrapolated per-particle Coulomb energies of the Laughlin and Pfaffian states in the lowest and second Landau level obtained using the spherical and disk pseudopotentials at various finite widths.

\begin{figure}[htpb!]
\begin{center}
\includegraphics[width=0.47\textwidth,height=0.29\textwidth]{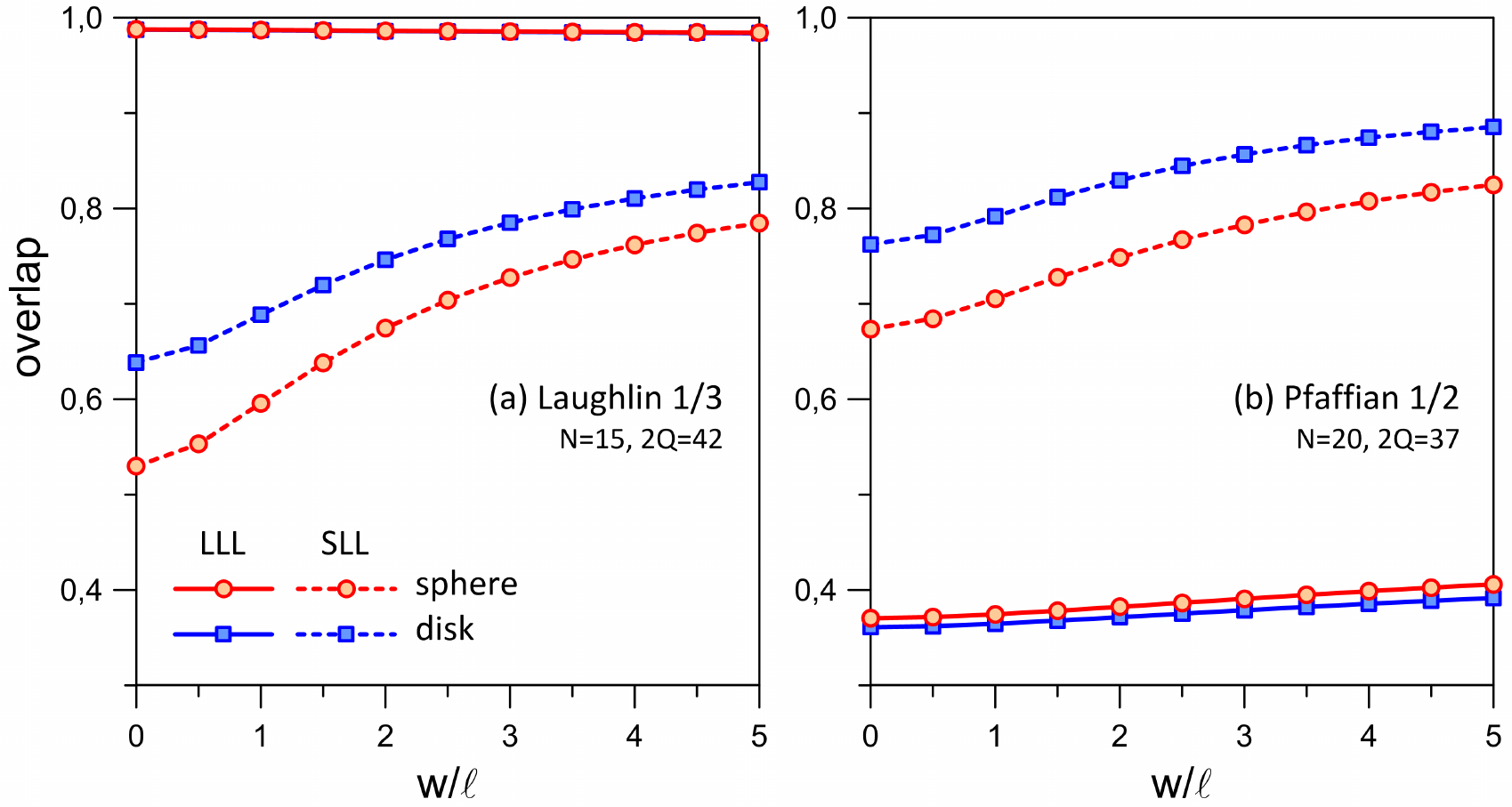} 
\caption{(color online) Overlap of the exact Laughlin state (left panel) and Pfaffian state (right panel) with the exact Coulomb ground state evaluated in the spherical geometry using the disk and spherical pseudopotentials for various widths $w$ for the largest system sizes considered in this work. }
\label{fig:overlaps_Laughlin_Pfaffian_sphere_disk}
\end{center}
\end{figure}

\begin{table*}[htpb!]
\centering
\begin{tabular}{|c|c|c|c|c|c|c|c|c|c|}
\hline 
$N$ & $2Q$ & $|\langle \psi_{1/3}^{\rm 1LL}|\psi_{1/3}^{L} \rangle|$ & $\langle \psi_{1/3}^{\rm 1LL}|H^{\rm 1LL}|\psi_{1/3}^{\rm 1LL}\rangle$ & $\langle \psi_{1/3}^{L}|H^{\rm 1LL}|\psi_{1/3}^{L}\rangle$ & $|\langle \psi_{1/3}^{\rm 0LL}|\psi_{1/3}^{L} \rangle|$ & $\langle \psi_{1/3}^{\rm 0LL}|H^{\rm 0LL}|\psi_{1/3}^{\rm 0LL}\rangle$ & $\langle \psi_{1/3}^{L}|H^{\rm 0LL}|\psi_{1/3}^{L}\rangle$ & $|\langle \psi_{1/3}^{\rm 1LL}|\psi_{1/3}^{\rm 0LL} \rangle|$\\ \hline
6	&	15	&	0.52848	&	-0.33306	&	-0.32604	&	0.99645	&	-0.41095	&	-0.41075	&	0.58814	\\ \hline
7	&	18	&	0.60705	&	-0.33017	&	-0.32611	&	0.99636	&	-0.41082	&	-0.41062	&	0.64941	\\ \hline
8	&	21	&	0.57197	&	-0.33071	&	-0.32615	&	0.99540	&	-0.41074	&	-0.41052	&	0.62836	\\ \hline
9	&	24	&	0.47941	&	-0.33101	&	-0.32617	&	0.99405	&	-0.41068	&	-0.41044	&	0.53871	\\ \hline
10	&	27	&	0.54000	&	-0.33043	&	-0.32619	&	0.99295	&	-0.41063	&	-0.41038	&	0.60592	\\ \hline
11	&	30	&	0.70303	&	-0.32954	&	-0.32620	&	0.99217	&	-0.41058	&	-0.41032	&	0.77057	\\ \hline
12	&	33	&	0.50296	&	-0.33039	&	-0.32621	&	0.99092	&	-0.41055	&	-0.41028	&	0.57650	\\ \hline
13	&	36	&	0.54445	&	-0.33001	&	-0.32622	&	0.98977	&	-0.41052	&	-0.41024	&	0.62202	\\ \hline
14	&	39	&	0.57706	&	-0.32984	&	-0.32623	&	0.98871	&	-0.41049	&	-0.41020	&	0.66098	\\ \hline
15	&	42	&	0.52978	&	-0.32991	&	-0.32623	&	0.98763	&	-0.41047	&	-0.41018	&	0.61475	\\ \hline
\end{tabular} 
\caption{\label{energies_overlaps_SLL_Laughlin} Table shows the overlap of the exact Coulomb ground state $|\psi_{1/3}^{\rm 1LL}\rangle$ for $N$ electrons at the Laughlin flux $2Q=3N-3$ in the second Landau level with the Laughlin state $|\psi_{1/3}^{L} \rangle$. We also show the per-particle density corrected second Landau level Coulomb energies (which includes the contribution of the positively charged background) of the exact and Laughlin states in Coulomb units of $e^{2}/(\epsilon\ell)$. For completeness, we have also included analogous numbers in the lowest Landau level. Some of these results were previously quoted in Refs.~\cite{Ambrumenil88,Balram13b,Kusmierz18}.}
\end{table*}

\begin{table*}[htpb!]
\centering
\begin{tabular}{|c|c|c|c|c|c|c|c|c|c|}
\hline 
$N$ & $2Q$ & $|\langle \psi_{1/2}^{\rm 1LL}|\psi_{1/2}^{\rm MR} \rangle|$ & $\langle \psi_{1/2}^{\rm 1LL}|H^{\rm 1LL}|\psi_{1/2}^{\rm 1LL}\rangle$ & $\langle \psi_{1/2}^{\rm MR}|H^{\rm 1LL}|\psi_{1/2}^{\rm MR}\rangle$ & $|\langle \psi_{1/2}^{\rm 0LL}|\psi_{1/2}^{\rm MR} \rangle|$ & $\langle \psi_{1/2}^{\rm 0LL}|H^{\rm 0LL}|\psi_{1/2}^{\rm 0LL}\rangle$ & $\langle \psi_{1/2}^{\rm MR}|H^{\rm 0LL}|\psi_{1/2}^{\rm MR}\rangle$ & $|\langle \psi_{1/2}^{\rm 1LL}|\psi_{1/2}^{\rm 0LL} \rangle|$\\ \hline
6$^{\dagger}$	&	9	&	0.81624	&	-0.37528	&	-0.37186	&	0.92280	&	-0.46734	&	-0.46275	&	0.53064	\\ \hline
8$^{\dagger}$	&	13	&	0.86739	&	-0.37195	&	-0.36966	&	0.92130	&	-0.46175	&	-0.45921	&	0.61067	\\ \hline
10	&	17	&	0.83764	&	-0.36968	&	-0.36795	&	0.90537	&	-0.46239	&	-0.45931	&	0.55247	\\ \hline
12$^{\dagger}$	&	21	&	0.81939	&	-0.36900	&	-0.36678	&	0.65520	&	-0.46651	&	-0.45922	&	0.24462	\\ \hline
14	$^{\dagger}$&	25	&	0.69345	&	-0.36850	&	-0.36621	&	0.72226	&	-0.46367	&	-0.45875	&	0.28091	\\ \hline
16	&	29	&	0.77945	&	-0.36767	&	-0.36574	&	0.74587	&	-0.46297	&	-0.45854	&	0.33699	\\ \hline
18	&	33	&	0.67656	&	-0.36764	&	-0.36536	&	0.63554	&	-0.46400	&	-0.45845	&	0.20746	\\ \hline
20$^{\dagger}$	&	37	&	0.67364	&	-0.36723	&	-0.36508	&	0.37029	&	-0.46621	&	-0.45834	&	0.08814	\\ \hline
\end{tabular} 
\caption{\label{energies_overlaps_SLL_MR_Pfaffian} Table shows the overlap of the exact Coulomb ground state $|\psi_{1/2}^{\rm 1LL}\rangle$ for $N$ electrons at the Moore-Read Pfaffian flux $2Q=2N-3$ in the second Landau level with the Moore-Read Pfaffian state $|\psi_{1/2}^{\rm MR} \rangle$. We also show the per-particle density corrected second Landau level Coulomb energies (which includes the contribution of the positively charged background) of the exact and Moore-Read Pfaffian states in Coulomb units of $e^{2}/(\epsilon\ell)$. For completeness, we have also included analogous numbers in the lowest Landau level, where overlaps and energies are shown for the \emph{lowest energy $L=0$ state} ($^{\dagger}$ indicates cases where the lowest energy $L=0$ state is also the ground state in the lowest Landau level). Some of these results were previously quoted in Refs.~\cite{Morf98,Scarola02,Pakrouski15,Kusmierz18}.}
\end{table*}

\subsection{Updating results of Ref.~\cite{Balram18a}}
The $\bar{3}\bar{2}111$ parton state, which is the predecessor of the $\bar{4}\bar{2}111$ parton state in the $\bar{n}\bar{2}111$ parton sequence, was posited as a candidate~\cite{Balram18a} to describe the experimentally observed FQHE $\nu=2+6/13$ FQHE~\cite{Kumar10}. In Ref.~\cite{Balram18a}, the $\bar{3}\bar{2}111$ parton state was compared against the exact SLL Coulomb ground state for only a single system of $N=12$ electrons at a flux of $2Q=28$. At the time of publication, the next system size, that of $N=18$ electrons at a flux of $2Q=41$, which has a Hilbert space dimension of $3.5\times 10^{9}$, was not accessible to exact diagonalization. We have now been able to evaluate the exact SLL Coulomb ground state for this system size, which, to the best of our knowledge, is the largest system size for which the ground state of an FQHE Hamiltonian has been obtained. Improvements in the matrix times vector multiplication at each Lanczos iteration, which have made it possible to efficiently diagonalize such large matrices, have been described in Ref.~\cite{Zuo20}.

\begin{figure}[htpb!]
\begin{center}
\includegraphics[width=0.47\textwidth,height=0.31\textwidth]{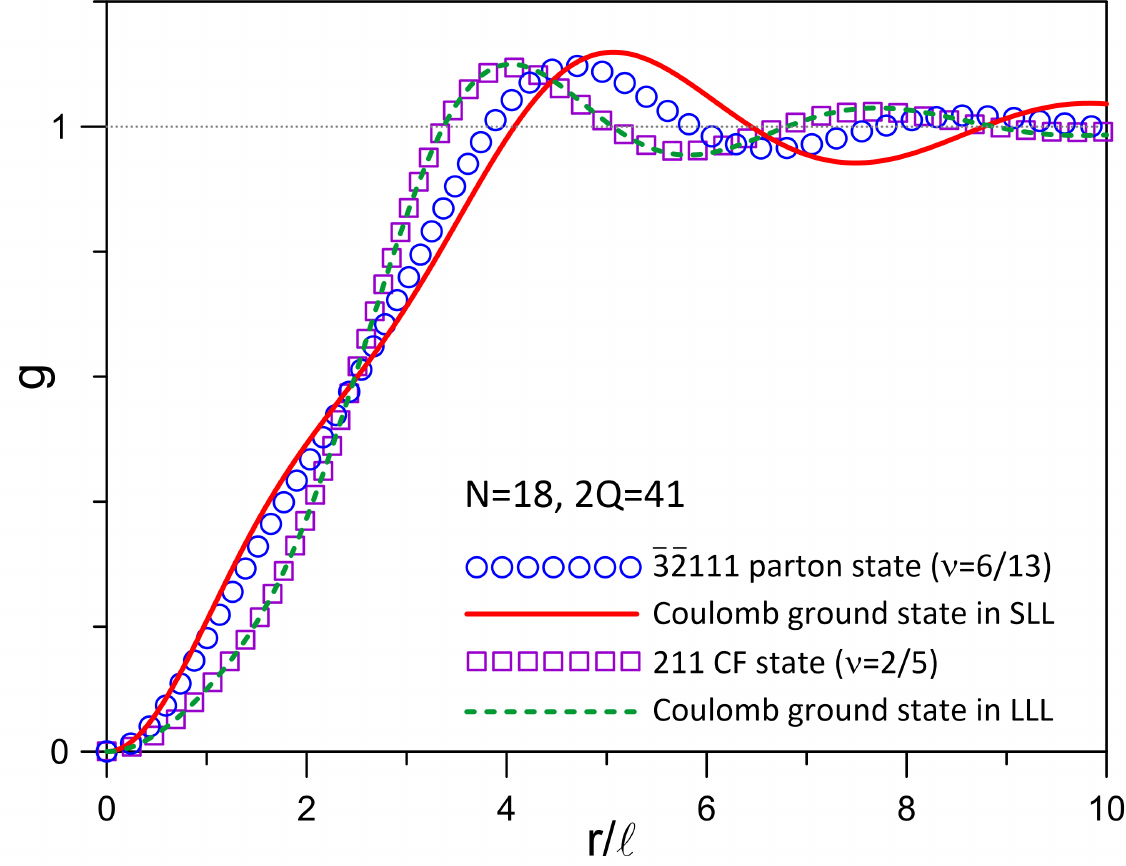} 
\caption{(color online) The pair correlation function $g(r)$ as a function of the arc distance for the exact second Landau level Coulomb ground state, and the $\bar{3}\bar{2}111$ parton candidate state for $N=18$ electrons at a flux of $2Q=41$. For comparison we also show $g(r)$ for the exact LLL Coulomb ground state as well as that of the $\nu=2/5$ Jain CF state for the same system size. }
\label{fig:pair_correlations_6_13}
\end{center}
\end{figure}

\begin{figure}[htpb!]
\begin{center}
\includegraphics[width=0.47\textwidth,height=0.31\textwidth]{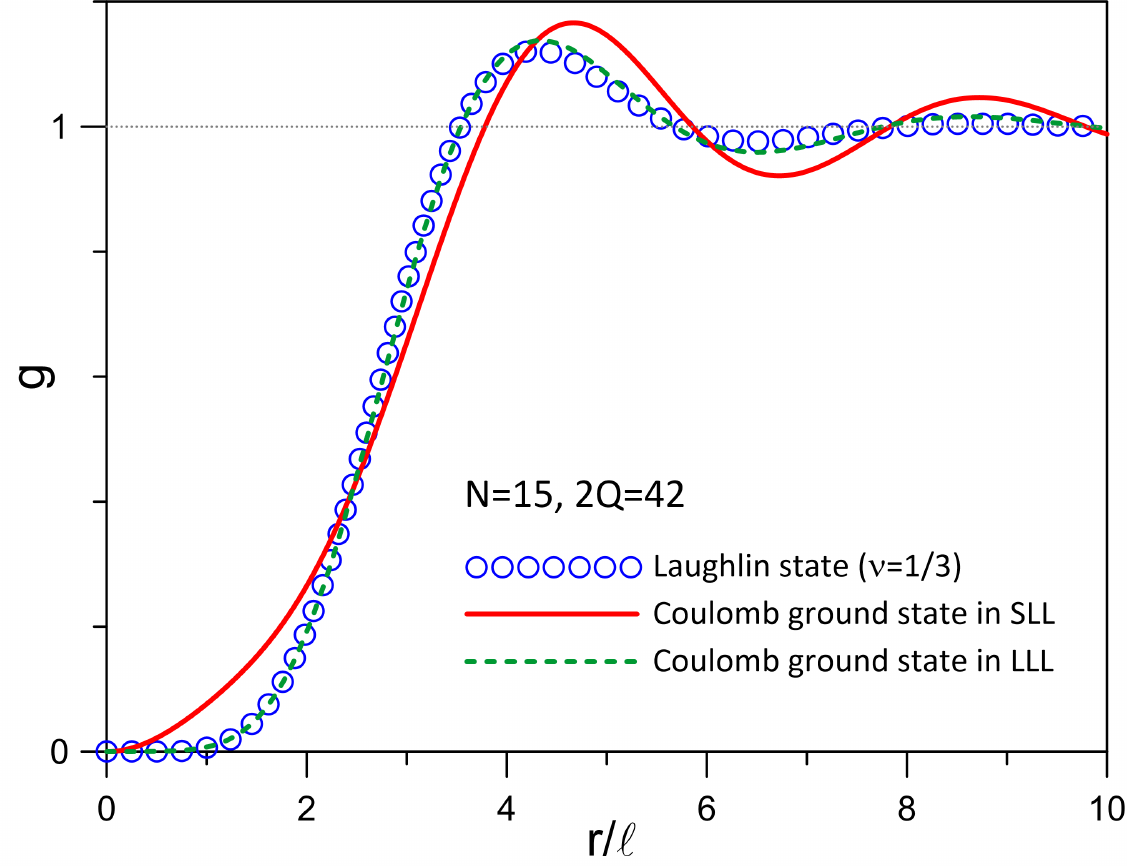} 
\caption{(color online) The pair correlation function $g(r)$ as a function of the arc distance for the exact second Landau level Coulomb ground state, and the Laughlin state for $N=15$ electrons at a flux of $2Q=42$. For comparision we also show $g(r)$ for the exact LLL Coulomb ground state for the same system. }
\label{fig:pair_correlations_1_3}
\end{center}
\end{figure}

\begin{figure}[htpb!]
\begin{center}
\includegraphics[width=0.47\textwidth,height=0.31\textwidth]{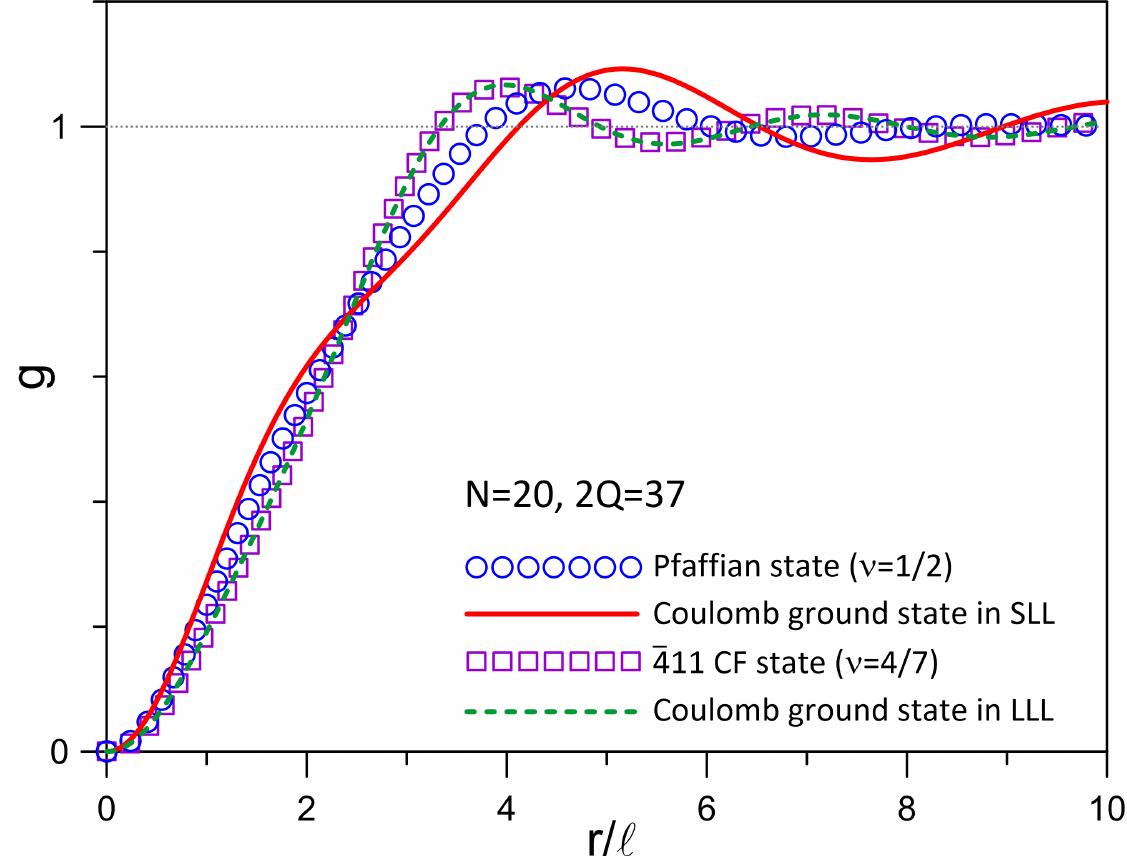} 
\caption{(color online) The pair correlation function $g(r)$ as a function of the arc distance for the exact second Landau level Coulomb ground state, and the Pfaffian state  for $N=20$ electrons at a flux of $2Q=37$. For comparison we also show $g(r)$ for the exact LLL Coulomb ground state as well as that of the $\nu=4/7$ Jain CF state for the same system size. }
\label{fig:pair_correlations_1_2}
\end{center}
\end{figure}

The exact SLL Coulomb ground state for $N=18$ electrons at a flux of $2Q=41$ has $L=0$ that is consistent with realizing a uniform state. This system has a neutral gap of 0.0096 $e^{2}/(\epsilon\ell)$. In Fig.~\ref{fig:pair_correlations_6_13} we compare the pair-correlation function of the exact SLL Coulomb ground state with that of the $\bar{3}\bar{2}111$ parton state. The pair-correlation functions of both the $\bar{3}\bar{2}111$ parton state as well as the exact SLL Coulomb ground state show oscillations that decay at long distances, which is a typical feature of an incompressible state~\cite{Kamilla97, Balram15b}. Furthermore, $g(r)$ of the exact SLL Coulomb ground state and the parton trial state match reasonably well with each other. Although the agreement is not conclusive, it is on par with that of other candidates in the SLL~\cite{Ambrumenil88, Morf98, Scarola02, Balram13b, Pakrouski15, Hutasoit16, Kusmierz18}. For example, in Figs.~\ref{fig:pair_correlations_1_3} and~\ref{fig:pair_correlations_1_2} we show a comparison of the $g(r)$'s of the Laughlin state~\cite{Laughlin83} with that of the $\nu=7/3$ ground state and of the Pfaffian state~\cite{Moore91} with that of the $\nu=5/2$ ground state for the largest systems considered in this work. Thus, we conclude that the results for the $N=18$ electrons at flux $2Q=41$ provide further evidence for the feasibility of the $\bar{3}\bar{2}111$ parton state to describe the $2+6/13$ FQHE.

Including the contribution of the positively charged background, the per-particle density corrected~\cite{Morf86} energy of the $\bar{3}\bar{2}111$ parton state for the effective interaction that we use to simulate the physics of the SLL in the LLL for the system of $N=18$ particles at flux $2Q=41$ is $-0.3855(1)$ while the exact energy of the ground state is $-0.3879$, both in Coulomb units of $e^{2}/(\epsilon\ell)$. Again, the level of agreement (within $0.64\%$) between these two numbers is comparable with that of other candidate wave functions in the second LL~\cite{Hutasoit16, Balram18a}. 

For completeness, we have also evaluated the exact LLL Coulomb ground state for the $N=18$ system at the flux of $2Q=41$. The overlaps of the LLL and SLL Coulomb ground states is $0.0872$ and their $g(r)$'s are also very different from each other (see Fig.~\ref{fig:pair_correlations_6_13}) which suggests that the ground state in the two LLs are quite different from each other for this system. We note that the system of $N=18$ electrons at a flux of $2Q=41$ aliases with the $\nu=2/5$ Jain CF state. As expected, we find excellent agreement between the $g(r)$'s of the LLL Coulomb ground state and the $\nu=2/5$ Jain CF state for this system (see Fig.~\ref{fig:pair_correlations_6_13}). Microscopically, the Jain CF states provide a very accurate representation of the exact LLL Coulomb ground states whereas the candidate states in the SLL provide only an approximate representation of the exact SLL Coulomb ground state. The neutral gap for the system of $N=18$ electrons at the flux $2Q=41$ in the LLL is 0.0487 $e^{2}/(\epsilon\ell)$ (five times larger than the corresponding SLL neutral gap quoted above) indicating that the 2/5 Jain CF state is more robust than the $6/13$ parton state.

\end{document}